\documentclass[lettersize,journal]{IEEEtran}
\usepackage{amsmath,amsfonts}
\usepackage{algorithmic}
\usepackage{algorithm}
\usepackage{array}
\usepackage[caption=false,font=normalsize,labelfont=sf,textfont=sf]{subfig}
\usepackage{textcomp}
\usepackage{stfloats}
\usepackage{url}
\usepackage{verbatim}
\usepackage{graphicx}
\usepackage{cite}
\usepackage{listings}
\usepackage{amsmath,amsfonts}
\usepackage{graphicx}
\usepackage{amsfonts}
\usepackage{array}
\usepackage{amssymb}
\usepackage{helvet}
\usepackage{color}
\usepackage{bm}
\usepackage{float}
\usepackage{cite}
\usepackage{multirow}
\usepackage{slashbox}
\usepackage{amssymb}
\usepackage[caption=false,font=normalsize,labelfont=sf,textfont=sf]{subfig}
\usepackage{amsmath,amsfonts}
\usepackage{algorithmic}
\usepackage{algorithm}
\usepackage{textcomp}
\usepackage{stfloats}
\usepackage{url}
\usepackage{verbatim}
\usepackage{graphicx}
\usepackage{cite}
\usepackage{amssymb}
\usepackage{makecell}

\newcommand{\RNum}[1]{\uppercase\expandafter{\romannumeral #1\relax}}

\begin{document}

\title{Beam Squint Assisted User Localization in Near-Field Integrated Sensing and Communications Systems}

\author{Hongliang Luo, Feifei Gao,~\IEEEmembership{Fellow,~IEEE,} Wanmai Yuan,~\IEEEmembership{Student Member,~IEEE,} \\ and Shun Zhang,~\IEEEmembership{Senior Member,~IEEE}
\thanks{
This paper has been accepted by IEEE Transactions on Wireless Communications (TWC) on 18 September 2023.
Manuscript received 2 September 2022; revised 20 May 2023 and 14 September 2023; accepted 18 September 2023; date of current version 24 September 2023.
The  editor  coordinating the review of this article and approving it for publication is Dr. Maxime Guillaud.
This work was supported in part  by the National Natural Science Foundation of China under Grant  {62325107, 62341107, 62261160650}, and by Beijing Natural Science Foundation under Grant L222002.   (\textit{Corresponding author: Feifei Gao.}) }
\thanks{H. Luo and F. Gao are with Institute for Artificial Intelligence, Tsinghua University (THUAI), State Key Lab of Intelligent Technologies and Systems, Tsinghua University,
Beijing National Research Center for Information Science and Technology (BNRist),
Department of Automation, Tsinghua University, 
Beijing, China (email: luohl23@mails.tsinghua.edu.cn, feifeigao@ieee.org).}
\thanks{
Wanmai Yuan is with Institute Information Science Academy of CETC (email: yuanwanmai7@163.com).}
\thanks{
Shun Zhang is  with the State Key Laboratory of Integrated Services Networks, Xidian University, China
(email: zhangshunsdu@xidian.edu.cn).}
}

\maketitle

\begin{abstract}
Integrated sensing and communication (ISAC) has been regarded as a key technology for 6G wireless communications, in which large-scale multiple input and multiple output (MIMO) array with higher and wider frequency bands will be adopted.
However, recent studies show that the beam squint phenomenon can not be ignored in  wideband   MIMO system, which generally deteriorates the communications performance. 
In this paper, we find that with the aid of  true-time-delay lines (TTDs), the range and trajectory of the beam squint in  near-field  communications systems can be freely controlled, and hence it is possible to reversely utilize the beam squint for user localization.
We derive the trajectory equation for \emph{near-field beam squint points} and design a way to control such trajectory. With the proposed design, beamforming from different subcarriers would purposely point to different angles and different distances, such that users from different positions would receive the maximum power at different subcarriers. 
Hence, one can simply localize multiple users from the beam squint effect in frequency domain, and thus reduce the beam sweeping  overhead as compared to the
conventional time domain beam search based approach.
Furthermore, we  utilize the phase difference of the maximum power subcarriers received by the user at different frequencies in several times beam sweeping to obtain a more accurate  distance estimation result, ultimately realizing high accuracy and low beam sweeping overhead user localization.
Simulation results demonstrate the effectiveness of the proposed schemes.
\end{abstract}


\section{Introduction}
ISAC has attracted much research interest recently and is deemed as one of the key technologies for  the sixth generation (6G)
 communications \cite{ISAC1,9737357,ISAC3,ISAC2}.
One of the ideas  is to use the communication signaling to sense the user's position as a simultaneous functionality. 
Once the user's position is obtained,  the base station (BS) system can not only  better serve the user's communication \cite{d2}, but also supports a variety of intelligent services, such as path planning for the Internet of Vehicles,
intrusion detection for intelligent security and unmanned aerial vehicle  cooperative work
 \cite{butalso1,jiankong,UAV1}.

When the user is far from the BS, say in the far-field area,
the BS typically  focuses on the users' angles in order to provide directional far-field communication beams in early research,
which can be regarded as the estimation of the direction of arrival (DOA) problem.
In the study of array signal processing, a number of classical DOA estimation methods were proposed. 
For example, multiple signal classification (MUSIC) algorithm can realize DOA estimation based on matrix eigenspace decomposition and is an important cornerstone of direction finding theory of spatial spectrum estimation \cite{music1}.
In addition, DOA estimation methods based on compressed sensing have also been proposed that take the advantage of the sparse characteristics of signals \cite{cs1,cs2,cs3}.
Recently, 
Z.~Chen  proposed a RIS-based  DOA estimation method for ISAC system \cite{DOAISACPZ}.
P.~Chen  proposed a super-resolution DOA estimation method based on deep learning, and discussed the basic problem of DOA estimation in ISAC system \cite{DOAISACPC}.

As  MIMO array becomes larger and larger in 5G and future 6G communications, 
the user's position may stay within the Rayleigh distance,
resulting in the evolution from far-field effect to near-field effect of electromagnetic field \cite{ruili,zhy11,thz1112}.
The Rayleigh distance  is usually expressed as $Z=\frac{2D^2}{\lambda}$, where  $\lambda$ is the wavelength and $D$ is antenna array aperture.
For example, the wavelength of a BS operating in  $ 30$ GHz  is 
$\lambda =0.01m$.  Assuming the BS is configured with a uniform linear array (ULA) with $128$ antenna  elements and half wavelength antenna spacing,
then the aperture of this BS is $D= \frac{(128-1)\times 0.01m}{2}=0.635m$, and
 the near-field Rayleigh distance of this BS is about $Z=\frac{2D^2}{\lambda}=\frac{2\times 0.635m^2}{0.01m}\approx 80m$. This distance is a typical working distance in the future smart city, smart transportation and smart factory \cite{near1}.
Thus we need to consider how to realize user localization in  near-field ISAC system.
There have been some interesting work in this area recently.
Y. Jiang  proposed a
reconfigurable intelligence surface
(RIS) assisted near-field communication and localization method \cite{9941256}.
J. Yang  proposed a localization method based on extremely large lens antenna array, which utilizes the window effect for energy focusing property of ExLens \cite{yang1}.

However, these methods all consider narrowband signals, but 6G communication will adopt  wider frequency bands as well as advanced beamforming  to achieve  higher rate communication transmission
\cite{sec1duan1cite1,sec1duan1cite2,sec1duan1cite3}.
It has been shown in \cite{wblsp}  that for a wideband communication system using orthogonal frequency division multiplexing (OFDM) modulation,
the phenomenon of \emph{beam squint} would appear in  far-field communications, in which the beamforming from different subcarriers would point to different directions, making the energy of part subcarriers deviate from the desired user position. 
Beam squint effect is usually considered as a negative effect and  should be mitigated with the aid of the true-time-delay lines (TTDs) \cite{b1,9896734,9839132}. 
For example, an efficient method was proposed to realize the wideband hybrid beamforming by using TTDs, which greatly alleviates the energy leakage caused by beam squint while considering the economic cost \cite{nn1}. 
Nevertheless, it is recently reported that  beam squint can be inversely used to sense the DOA for different users thanks to the squint beam over different subcarriers \cite{10058989}.

When the user moves to  near-field area, the beam squint phenomenon will also occur.
Unlike the plane wave assumption for the far-field, the spherical wave assumption is usually used for the near-field \cite{waveassume1,waveassume2,waveassume3}. 
In this case, due to the propagation characteristics of the spherical waves,
beamforming from different subcarriers will focus at different positions.
This property is conducive for us to locate the positions of near-field users,
but 
there is no related discussions in literature, to the best of the authors' knowledge.

In this paper, we fill this research gap and propose the low-overhead near-field user localization algorithms by reversely utilizing the wideband beam squint effect, in which the BS can use OFDM signals to quickly sense the users' positions in frequency domain. 
The contributions of this paper are summarized as follows.

\begin{itemize}

\item We derive the near-field wideband channel models in  time domain and frequency domain as well as the corresponding beamforming models.

\item We analyze the near-field beam squint phenomenon and derive the trajectory equation of \emph{near-field beam squint points}. Then we design a way to control the range and trajectory of the beam squint points with the help of TTDs.

\item We reversely utilize the controllable beam squint for user localization and 
 propose a low complexity near-field user localization scheme based on TTDs-assisted beam squint.

\item We  propose a high accuracy localization scheme based on controllable beam squint, which utilizes the phase difference of   maximum power subcarriers with different frequencies received by users in several times beam sweeping to realize more accurate distance estimation.

\end{itemize}

The remainder of this paper is organized as follows. 
In Section \RNum{2}, we introduce the system model of the near-field communications.
In Section \RNum{3}, the near-field beam squint phenomenon is analyzed
and we propose a controllable beam squint strategy with the help of TTDs.
Then we proposed several  near-field user localization schemes based on controllable beam squint  in Section \RNum{4}. 
Simulation results and conclusions are given in Section \RNum{5} and Section \RNum{6}.

\emph{Notations}:
Lower-case and upper-case boldface letters $\mathbf{a}$ and $\mathbf{A}$ denote a vector and a matrix, respectively;
$\mathbf{a}^T$ and $\mathbf{a}^H$ denote the transpose and the conjugate transpose of vector $\mathbf{a}$, respectively; 
${\rm diag}(\mathbf{a})$ denotes a diagonal matrix with the diagonal elements constructed from $\mathbf{a}$.
$[\mathbf{a}]_n$  denotes the $n$-th element of the vector $\mathbf{a}$.
Symbols $\textbf{1}$ represent the all-ones matrix or vector.
$\mathcal{R}\{\cdot\}$ and $\arg(\cdot)$ denote the real part and the phase of a complex number, a vector or a matrix, respectively.
$\lceil x \rceil$ denotes the minimum integer that is not smaller than $x$.
$\delta(\cdot)$ is the Dirac delta function, and
$\left|\cdot\right|$  denotes the absolute operator.

\section{System Model}
A wideband massive MIMO system
working in mmWave frequency band with OFDM modulation is considered in this paper, in which
the BS is equipped with an $N$-antenna ULA with antenna spacing $d$ and a single radio frequency (RF) chain. 
The $n$-th antenna is deployed  at  $(0, nd)$, where 
$n=-\frac{N-1}{2}, ...,\frac{N-1}{2}$,
and the antenna array aperture is $D=(N-1)d\approx Nd$.
The carrier frequency and transmission bandwidth are denoted as ${f}_c$ and $W$,
and the subcarrier frequency range is $[{f}_c-\frac{W}{2},{f}_c+\frac{W}{2}]$.
Assuming there are a total of $M+1$ subcarriers, 
the $0$-th subcarrier has the lowest frequency $f_0={f}_c-\frac{W}{2}$
and  the $m$-th subcarrier frequency is $f_m={f}_0+m\frac{W}{M}$.
Particularly, by denoting the baseband frequency of the $m$-th subcarrier as
$\widetilde{f}_m=m\frac{W}{M}$, 
there is  $f_m=f_0+\widetilde{f}_m$, where $m=0,1,2,...,M$.
Assume that the BS serves $K$ users, and each user is equipped with a single antenna.
Suppose that the $k$-th user is located at $(x_k,y_k)$, and the corresponding polar coordinate is $(r_k,\theta_k)$.
Then the distance between the $n$-th antenna and the $k$-th user is $r_{k,n}=\sqrt{x_k^2+(y_k-nd)^2}$.

\begin{figure}
\centering
\includegraphics[width=70mm]{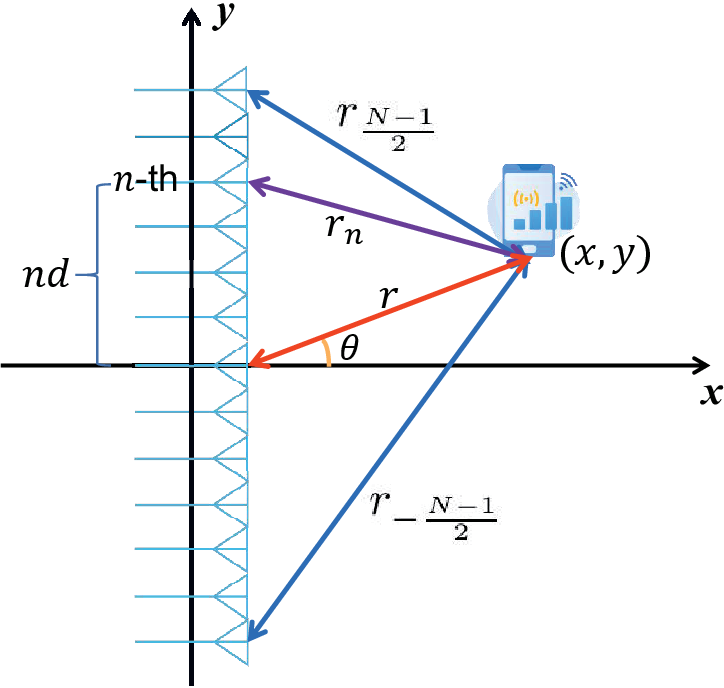}
\caption{Near-field system model.}
\label{fig1:env}
\end{figure}

According to \cite{ruili}, the boundary between  near-field and far-field is determined by the Rayleigh distance $Z=\frac{2D^2}{\lambda}$, where  $\lambda$ is the wavelength.
Here all users are assumed to be in the near-field range.
For mmWave frequency band communications, we consider only one line-of-sight (LoS) channel between the user and the BS while the proposed study can be readily extended to the multipath scenario. 
Denote $\tau_{k,n}$ as the time delay of the path from
the $n$-th antenna of the BS to  the antenna of the $k$-th user. 
Under the near-field assumption, there is $\tau_{k,n}=\frac{r_{k,n}}{c}$,
where $c$ represents the speed of light.
In the near-field area, the Fresnel approximation \cite{jinsi} is usually employed to expand $r_{k,n}$ as
\begin{equation}
\begin{split}
\begin{aligned}
\label{deqn_ex1a}
r_{k,n}&=\sqrt{x_k^2+(y_k-nd)^2}=\sqrt{r_k^2+n^2d^2-2r_knd\sin\theta_k}\\
&\approx r_k\!\!\left[1\!+\!\frac{1}{2}(\frac{n^2d^2}{r_k^2}\!-\!\frac{2nd\sin\theta_k}{r_k})\!-\!\frac{1}{8}(\frac{n^2d^2}{r_k^2}\!-\!\frac{2nd\sin\theta_k}{r_k})^2\right]\\
&\approx r_k\!-\!nd\sin\theta_k+\frac{n^2d^2\cos^2\theta_k}{2r_k}.
\end{aligned}
\end{split}
\end{equation}

Suppose that the signal transmitted by the BS is $s(t)$, and
the corresponding passband signal can be represented as $\mathcal{R}\{s(t)e^{j2\pi f_0t}\}$.
Then the signal received by the $k$-th user will be a delayed version of the transmit signal with amplitude attenuation. 
Specifically, the signal that the $k$-th user receives from the $n$-th antenna is 
$\mathcal{R}\left\{\alpha_{k,n} s(t-\tau_{k,n})e^{j2\pi f_0(t-\tau_{k,n})}\right\}$, where $\alpha_{k,n}$ is the channel path    loss.
Since it is a downlink channel, the $k$-th user eventually receives the signal as 
$\mathcal{R}\left\{\sum_{n=-\frac{N-1}{2}}^{\frac{N-1}{2}}\alpha_{k,n} s(t-\tau_{k,n})e^{j2\pi f_0(t-\tau_{k,n})}\right\}$.
Then the time-domain downlink channel between the $n$-th antenna of the BS
and  the $k$-th user 
can be expressed as
\begin{equation}
\label{deqn_ex1a}
h_{k,n}(t)=\alpha_{k,n} e^{-j2\pi f_0\tau_{k,n}}\delta(t-\tau_{k,n}),
\end{equation}
where $\delta(\cdot)$ denotes the Dirac delta function. 
By taking the Fourier transform of (2), 
the corresponding frequency-domain channel on the  $m$-th subcarrier can be represented as
\begin{equation}
\label{deqn_ex1a}
h_{k,n,m}^F =\alpha_{k,n,m} e^{-j2\pi f_m\tau_{k,n}}.
\end{equation}

According to  \cite{xindao}, the specific channel fading can be modeled as $\alpha_{k,n,m}=\frac{c}{4\pi f_m r_{k,n}}$.
Because the distance between the user and the center of  BS  is generally much larger than the antenna array aperture, i.e., $r_k \gg D$, it is usually assumed that
$\alpha_{k,-(N-1)/2,m}\approx...\approx\alpha_{k,(N-1)/2,m}\approx \alpha_{k,m} = \frac{c}{4\pi f_m r_k}$.
Then the LoS channel vector of the $k$-th user $\mathbf{h}(r_k,\theta_k,{f}_m)\in \mathbb{C}^{N\times 1}$ can be  modeled as
\begin{equation}
\begin{split}
\begin{aligned}
\label{deqn_ex1a}
[\mathbf{h}(r_k,\theta_k,{f}_m)]_n&=\alpha_{k,m}e^{-j2\pi f_m\frac{r_{k,n}}{c}}\\&=
\alpha_{k,m}e^{-j2\pi f_m\frac{r_k-nd\sin\theta_k+\frac{n^2d^2\cos^2\theta_k}{2r_k}}{c}},
\end{aligned}
\end{split}
\end{equation}
where $[\mathbf{h}(r_k,\theta_k,{f}_m)]_n$ denotes 
 the channel between the $n$-th antenna and the $k$-th user on the
$m$-th subcarrier.

In order to provide better communication links for the users, BS adopts the phase-shifters (PSs) based beamforming for the single RF chain.
We consider that the optimal near-field
 beamforming vector  $\mathbf{w}(r_0,\theta_0)\in \mathbb{C}^{N\times 1}$ is determined by the point $(r_0,\theta_0)$ and the lowest carrier frequency $f_0$,
that is,
 $\mathbf{w}(r_0,\theta_0)$ should match the phase of $\mathbf{h}(r_0,\theta_0,f_0)$ and can be denoted as
\begin{equation}
\label{deqn_ex1a}
[\mathbf{w}(r_0,\theta_0)]_n=\frac{1}{\sqrt{N}}e^{j\arg([\mathbf{h}(r_0,\theta_0,f_0)]_n)}=\frac{1}{\sqrt{N}}e^{-j2\pi f_0\frac{r_{0,n}}{c}},
\end{equation}
where $\arg(x)$ denotes the phase of $x$.
Then the downlink transmission process of the $k$-th user on the $m$-th subcarrier can be expressed as
\begin{equation}
\label{deqn_ex1a}
y_{k,m}=\mathbf{h}_{k,m}^H\cdot \mathbf{w}\cdot  s_{k,m}+n_{k,m},
\end{equation}
where $y_{k,m} \in \mathbb{C}^{1\times 1}$ denotes the complex signal received by the $k$-th user on the $m$-th subcarrier.
Moreover, $\mathbf{h}_{k,m}=\mathbf{h}(r_k,\theta_k,{f}_m)$ is the channel vector,
$s_{k,m}$ is the data transmitted by the $m$-th subcarrier, and $n_{k,m}$ is the zero-mean additive complex Gaussian white noise with variance of $\sigma_{k}^2$.
Then the complex signal received by the $k$-th user at each subcarrier can  be formed into a vector $\mathbf{y}_k\in \mathbb{C}^{(M+1)\times 1}$ as
\begin{equation}
\begin{split}
\begin{aligned}
\label{deqn_ex1a}
\mathbf{y}_k=[y_{k,0}, y_{k,1},...,y_{k,M}]^T={\rm diag}(\mathbf{H}_k^H\cdot\mathbf{w})\cdot\mathbf{s}_{k}+\mathbf{n}_{k},
\end{aligned}
\end{split}
\end{equation}
where $\mathbf{H}_k\triangleq[\mathbf{h}_{k,0},\mathbf{h}_{k,1},...,\mathbf{h}_{k,M}]\in \mathbb{C}^{N\times (M+1)}$ represents the overall downlink channel at different subcarriers in  frequency domain,
$\mathbf{s}_{k}\triangleq[s_{k,0},s_{k,1},...,s_{k,M}]^T\in \mathbb{C}^{(M+1)\times 1}$ represents the data information carried by all subcarriers, and $\mathbf{n}_{k}\triangleq[n_{k,0},n_{k,1},...,n_{k,M}]^T\in \mathbb{C}^{(M+1)\times 1}$ represents the additive noise vector.

\section{Near-Field Beam Squint}

\subsection{BS View: Trajectory of Near-Field Beam Squint Points}

In order to concentrate the beam transmitted by the BS at the desired near-field position $(r_0,\theta_0)$, under the limitations of only one single RF and the PSs structure, the BS needs to employ the near-field beamforming vector $\mathbf{w}(r_0,\theta_0)$ for all the subcarriers.
In narrowband systems, due to the small frequency difference between different subcarriers, all the subcarriers will be approximately focused at the desired position $(r_0,\theta_0)$.
However,
with wideband communications, i.e., $W$ is large, the beamforming of the $m$-th subcarrier may not point to the desired position $(r_0,\theta_0)$. Such a phenomenon is named as \emph{near-field beam squint} as the near-field beamforming position gradually ``squint'' over the frequency.

Specifically,
based on formula (4), the near-field array response vector at any  position $(r,\theta)$ on the $m$-th subcarrier can defined as $\mathbf{b}(r,\theta,{f}_m)\in \mathbb{C}^{N\times 1}$, which satisfies $[\mathbf{b}(r,\theta,{f}_m)]_n = e^{-j2\pi f_m\frac{r_{n}}{c}}$. Moreover, $r_n=\sqrt{x^2+(y-nd)^2}$ denotes the distance between the near-field point $(r,\theta)$ and the $n$-th  antenna of BS.
When $\mathbf{w}(r_0,\theta_0)$ is designed by the  desired position $(r_0,\theta_0)$ and frequency $f_0$,  the array gain at any position $(r, \theta)$ on the $m$-th subcarrier can be calculated as 
\begin{equation}
\begin{split}
\begin{aligned}
\label{deqn_ex1a}
g(r,\theta,{f}_m,r_0,\theta_0)\!&=\!\left| \mathbf{w}^H(r_0,\theta_0)\!\cdot\!\mathbf{b}(r,\theta,{f}_m)\right|\!\\&=\!\frac{1}{\sqrt{N}}\left| \sum _{n=-\frac{N-1}{2}}^{\frac{N-1}{2}} \!\!\!\!\!
e^{j2\pi f_0\frac{r_{0,n}}{c}}e^{-j2\pi f_m\frac{r_n}{c}}
\right|,
\end{aligned}
\end{split}
\end{equation}
where $r_{0,n}=\sqrt{x_0^2+(y_0-nd)^2}$ denotes the distance between the desired point $(r_0,\theta_0)$ and the $n$-th  antenna of BS.
Then we can apply
the Fresnel approximation as (1) for both  $r_{0,n}$ and   $r_{n}$ in (8) and obtain 
\begin{equation}
\begin{split}
\begin{aligned}
\label{deqn_ex1a}
&\!\!\!\!g(r,\theta,{f}_m,r_0,\theta_0)\!\!\!\!\\&\!\!=\!\!\frac{1}{\sqrt{N}}\!\!\left| \!\sum _{n=-\!\frac{N\!-\!1}{2}}^{\frac{N-1}{2}} \!\!\!\!\!\!
e^{j\!\frac{2\pi n^2d^2}{c}\!(f_m\!\frac{\cos^2\theta}{2r}\!-\!f_0\!\frac{\cos^2\theta_0}{2r_0})}\!
e^{-j\!\frac{2\pi nd}{c} (f_m\!\sin\theta\!-\!f_0\sin\theta_0)}
\right|.
\end{aligned}
\end{split}
\end{equation}

The maximum value of the array gain $g(x,y,{f}_m,x_0,y_0)$ is $\sqrt{N}$.
When $m=0$, ${g}$ is obviously maximized at the position $(r,\theta)=(r_0,\theta_0)$.
However, for the other subcarriers with frequency $f_m$,  $m\neq0$, the position
$(r,\theta)$ that maximizes ${g}$ is not $(r_0,\theta_0)$, and will be defined as $(r_m,\theta_m)$,
i.e., the beamforming focus squint to other position. 
We call $(r_m,\theta_m)$ as \emph{near-field beam squint points} in this paper, which satisfies
\begin{equation}
\begin{split}
\begin{aligned}
\label{deqn_ex1a}
&(r_m,\theta_m)\\&= \mathop{\mathrm{argmax}}\limits_{(r,\theta)}{\frac{1}{\sqrt{N}}\left| \sum _{n=-\frac{N-1}{2}}^{\frac{N-1}{2}} 
e^{j\frac{2\pi n^2d^2}{c}\chi_{1,m}}
e^{-j\frac{2\pi nd}{c}\chi_{2,m}}
\right|}\\
&=\mathop{\mathrm{argmax}}\limits_{(r,\theta)}\frac{1}{\sqrt{N}}\
F\left(\frac{2\pi d^2}{c}\chi_{1,m}, \frac{-2\pi d}{c}\chi_{2,m}\right),
\end{aligned}
\end{split}
\end{equation}
where $\chi_{1,m}=f_m\frac{\cos^2\theta}{2r}-f_0\frac{\cos^2\theta_0}{2r_0}$, 
 $\chi_{2,m}=f_m\sin\theta-f_0\sin\theta_0$, and 
 the binary function $F(x,y)$ is defined as $F(x,y) = |\sum_{n=-(N-1)/2}^{(N-1)/2}e^{jn^2x}e^{jny}|$.
Since 
$F\left(\frac{2\pi d^2}{c}(f_m\frac{\cos^2\theta}{2r}-f_0\frac{\cos^2\theta_0}{2r_0}), \frac{-2\pi d}{c}(f_m\sin\theta-f_0\sin\theta_0)\right) \leq N$ and $F(0,0)=N$,
considering the finite range of the near-field, a feasible solution for $(r_m,\theta_m)$ is
\begin{align}
& \sin\theta_m =\frac{f_0}{f_m}\sin\theta_0,\label{1}\\
& r_m =r_0\cdot\frac{f_m}{f_0}\cdot\frac{\cos^2\theta_m}{\cos^2\theta_0}\label{2},
\end{align}
where $\cos^2\theta_m=1-\sin^2\theta_m=1-(\frac{f_0}{f_m}\sin\theta_0)^2$. 

\begin{figure}[]
\centering
\includegraphics[width=88mm]{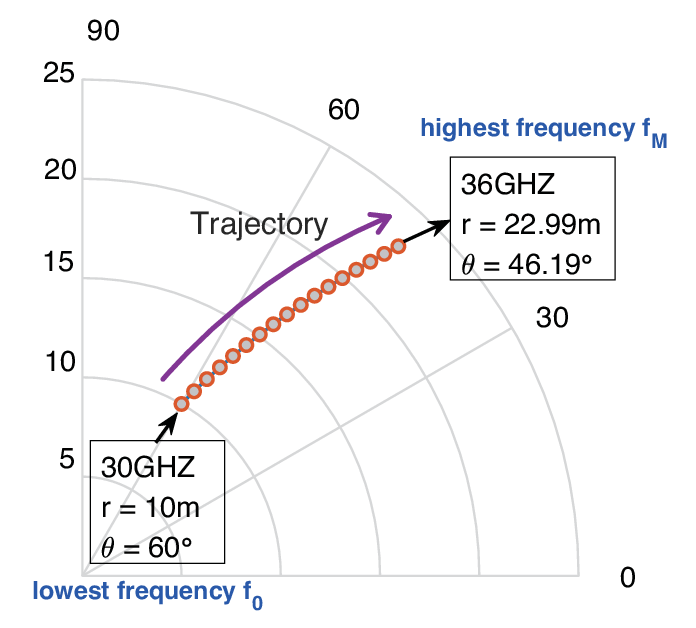}
\caption{Trajectory of near-field beam squint points, where $f_0=$ 30 GHz, $W=$ 6 GHz, $M=$16, the beamforming of the subcarrier with the lowest frequency  $f_0$ is set to focus on $(10m, 60^{\circ}).$}
\label{fig_1}
\end{figure}

Hence the angle and distance of the near-field beam squint point  of the $m$-th subcarrier is determined by  (11) and (12), respectively.
An example is presented in Fig.~2, which clearly shows the case of the near-field beam squint  from the BS view.
It is seen that
as the subcarrier frequency $f_m$ increases, the beamforming from different subcarriers would focus on different positions and can be connected into a curve trajectory.
The start point of the trajectory is determined by the $0$-th subcarrier with frequency ${f}_0$ while the end point of the trajectory is determined by the $M$-th subcarrier with frequency ${f}_M$.
In Fig.~2, the lowest frequency is $f_0=30$ GHz whose beamforming focuses at 
$(10m, 60^{\circ})$, while the highest frequency subcarrier is $36$ GHz whose beamforming eventually
squints to $(22.99m, 46.19^{\circ})$.
Clearly, near-field beam squint cannot be ignored in the wideband ISAC system.

\subsection{Users View: Power Spectrum of All Subcarriers}

As shown in formula (7), for the $k$-th user, the complex received signal at different subcarriers can constitute a vector $\mathbf{y}_k$. 
By taking the modulus  of each element in $\mathbf{y}_k$, the power of the $k$-th user's received signal at different frequencies can be obtained as
\begin{equation}
\begin{split}
\begin{aligned}
\label{deqn_ex1a}
\mathbf{g}_k=\left[|y_{k,0}|,|y_{k,1}|,...,|y_{k,M}|\right]^T,
\end{aligned}
\end{split}
\end{equation}
where the specific expression for $y_{k,m}$ is given by  (6).  Taking the index of the subcarrier $m$ as the horizontal axis, 
and taking the value of $|y_{k,m}|$ as the vertical axis,
we can draw the curve of the power spectrum of the $k$-th user's received signal.

\begin{figure}
\centering
\includegraphics[width=90mm]{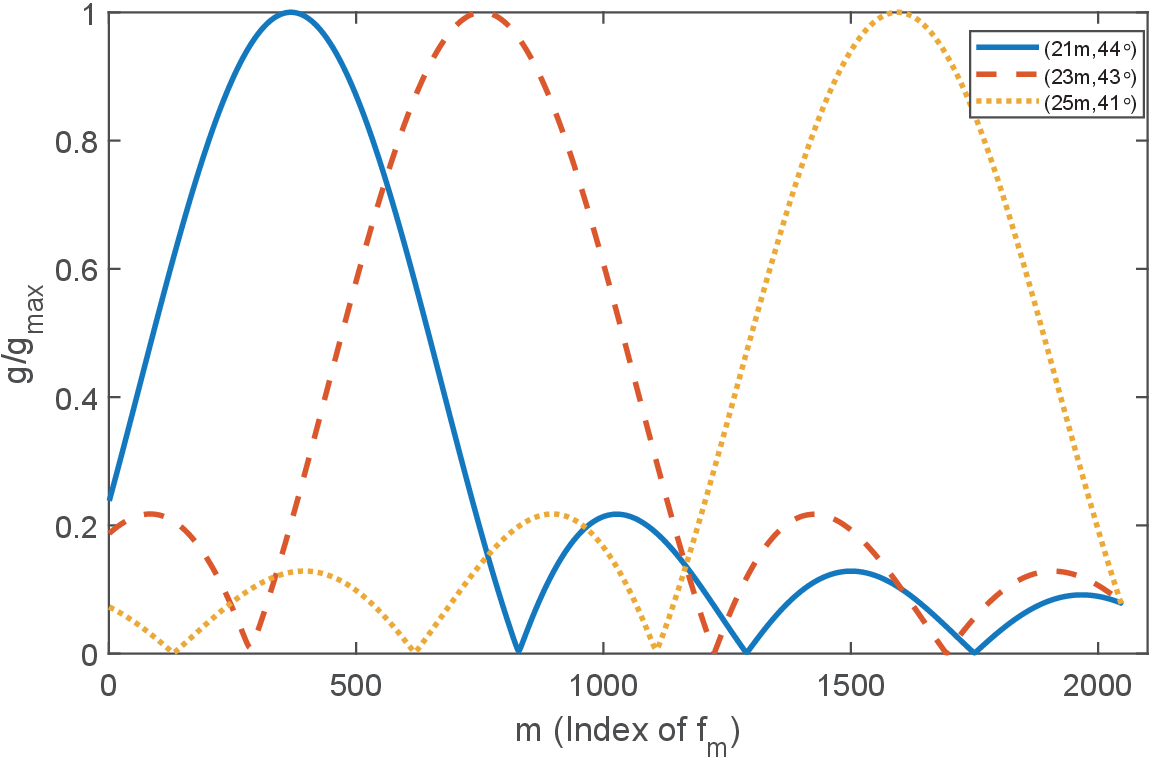}
\caption{The normalized power spectrums of received signals of different users under natural beam squint, where the users are located at $(21m,44^\circ)$, $(23m,43^\circ)$ and $(25m,41^\circ)$.}
\label{fig1:env}
\end{figure}

An example is shown in Fig.~3, in which the lowest frequency $f_0$ is 30 GHz, the highest frequency $f_M$ is 33 GHz, and the number of subcarriers is $M+1=2048$ . By adjusting the PSs such that the beamforming of
 the lowest frequency subcarrier  points to $(20m, 45^\circ)$, then the beamforming of the highest frequency subcarrier will squint to $(25.82m,40.00^\circ)$. 
We plot the received power spectrum  of three different users located within
the range of
 $\{(r,\theta)|20m\leq r \leq 25.82m, 40^\circ \leq \theta \leq 45^\circ\}$
 in Fig.~3. 
It is clearly seen that the power spectrums of the users at different positions are different, 
which will help BS to sense the locations of the users according to their special power spectrums.
By doing this, one may reduce the time of beam sweeping as compared to the narrowband system, i.e., the beam squint effect surprisingly has the positive effect when sensing the users' positions. 
Nevertheless, 
the  range of near-field beam squint  mainly depends on the relative transmission bandwidth of the system, and it is difficult to cover the entire required sensing range within only one  OFDM symbol.
Therefore, we must consider further expanding the range of beam squint to realize efficient sensing.

\subsection{BS View: Controllable Beam Squint  Based on TTDs}

Different from a PS that generates a fixed phase over the whole frequency band, a TTD line can provide a programmable true time delay that induces the varying phase over frequencies \cite{7394105,radarTTDs2}.
As shown in Fig.~4, 
we assume that each phase shifter is cascaded with one TTD line\cite{10058989}.
 The response of  the $n$-th phase shifter can be expressed  as $e^{-j2\pi\phi_n}$,
where $\phi_n$ denotes the phase shift amount of the $n$-th PS.
The time domain response of 
the $n$-th TTD can be expressed as $\delta(t-t_n)$, 
and the corresponding frequency domain response is $e^{-j2\pi \widetilde{f}t_n}$,
where
$t_n$ denotes the time delay amount of the $n$-th TTD and $\widetilde{f}$ is the baseband frequency.
Hence the new array beamforming vector assisted by TTDs can be expressed as
$\mathbf{\tilde{w}} \in \mathbb{C}^{N\times 1}$, and
\begin{equation} 
\begin{split}
\begin{aligned}
\label{deqn_ex1a}
[\mathbf{\tilde{w}}]_n=\frac{1}{\sqrt{N}}e^{-j2\pi\phi_n}e^{-j2\pi \widetilde{f}t_n}.
\end{aligned}
\end{split}
\end{equation}

\begin{figure}[]
\centering
\includegraphics[width=88mm]{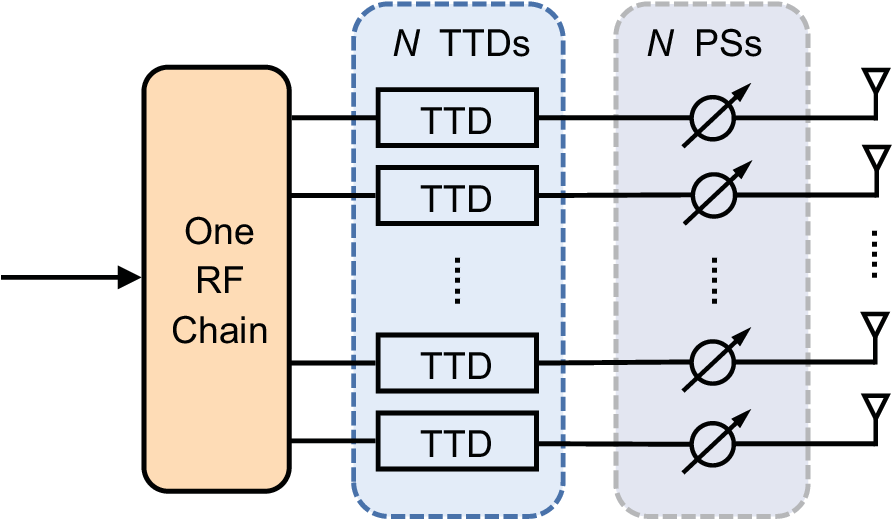}
\caption{Schematic diagram of the proposed TTDs assisted BS link.}
\label{fig_1}
\end{figure}

With TTDs, the array gain at any near-field position $(r, \theta)$ on the $m$-th subcarrier is
\begin{equation}
\begin{split}
\begin{aligned}
\label{deqn_ex1a}
\!\!\!\!\tilde{g}(r,\theta,f_m,\mathbf{\tilde{w}}) &=\left| \mathbf{\tilde{w}}^H \cdot \mathbf{b}(r,\theta,{f}_m)\right|\\&=\!\!\frac{1}{\sqrt{N}}\left| \sum _{n=-\frac{N-1}{2}}^{\frac{N-1}{2}} \!\!\!\!\!\!\!
e^{j2\pi\phi_n}e^{j2\pi \widetilde{f}_mt_n}e^{-\!j2\pi f_m\frac{r_{n}}{c}}
\right|.
\end{aligned}
\end{split}
\end{equation}

Similarly, $\tilde{g}(r,\theta,f_m,\mathbf{\tilde{w}})\leq \sqrt{N}$ and one of the solutions to maximize $\tilde{g}(r,\theta,f_m,\mathbf{\tilde{w}})$ 
need the following equation to be satisfied
\begin{equation}
\begin{split}
\begin{aligned}
\label{deqn_ex1a}
f_m\frac{r_{n}}{c}-\phi_n-\widetilde{f}_mt_n=p_{m,n},
\end{aligned}
\end{split}
\end{equation}
where $p_{m,n}$ is a set of integers introduced to align the phases in (15), and 
 equation (16) should hold for all the $n=-\frac{N-1}{2},...,\frac{N-1}{2}$.
In addition, we mark the near-field beam squint point of the $m$-th subcarrier assisted by TTDs as $(\tilde{r}_m,\tilde{\theta}_m)$, i.e.,
\begin{equation}
\begin{split}
\begin{aligned}
\label{deqn_ex1a}
\!\!\!\!(\tilde{r}_m,\!\tilde{\theta}_m)\!\!=\! \mathop{\mathrm{argmax}}\limits_{(r,\theta)}
\!\frac{1}{\sqrt{N}}\!\left| \!\sum _{n=-\frac{N\!-\!1}{2}}^{\frac{N-1}{2}} \!\!\!\!\!\!
e^{j2\pi\phi_n}\!e^{j2\pi \widetilde{f}_m\!t_n}\!e^{-\!j2\pi f_m\!\frac{r_{n}}{c}}\!
\right|.
\end{aligned}
\end{split}
\end{equation}

Although the start point and the end point of the beam squint trajectory are determined by subcarrier $f_0$ and $f_M$ respectively,
we can control the start point and the end point of the beam squint trajectory with the help of TTDs.
By adjusting the values of  the PSs, 
we can let the beamforming of the $0$-th subcarrier with frequency $f_0$ point to the start point $(r_0,\theta_0)$.
Specifically,
we set ${f}_m={f}_0$, $\widetilde{f}_m=\widetilde{f}_0=0$, $r_n=r_{0,n}$, $p_{m,n}=0$ in (16), and derive $\phi_{n}=\frac{f_0r_{0,n}}{c}$.
Then we set the phase shift of the $n$-th PS as $\phi_{n}$.
By adjusting the values of the TTDs, 
we can let the beamforming of the $M$-th subcarrier with frequency $f_M$ point to the end point $(r_c,\theta_c)$.
Specifically,
we set ${f}_m={f}_M$, $\widetilde{f}_m=\widetilde{f}_M=W$, $r_n=r_{c,n}$, $p_{m,n}=0$ in (16), and derive
$t_{n}=\frac{f_M}{Wc}r_{c,n}-\frac{\phi_{n}}{W}$. Then we set the time delay value of the $n$-th TTD as $t_{n}$.

Afterward, the beams from the $0$-th subcarrier to the $M$-th subcarrier will squint from the start point $(r_0,\theta_0)$ to the end point $(r_c,\theta_c)$. 
By substituting $\phi_{n}=\frac{f_0r_{0,n}}{c}$ and $t_{n}=\frac{f_M}{Wc}r_{c,n}-\frac{\phi_{n}}{W}$ in (17), the \emph{near-field controllable beam squint point}  of the $m$-th subcarrier  $(\tilde{r}_m,\tilde{\theta}_m)$
 satisfies 
\begin{equation}
\begin{split}
\begin{aligned}
\label{deqn_ex1a}
\!\!\!\!(\tilde{r}_m,\!\tilde{\theta}_m) \!&=\! \mathop{\mathrm{argmax}}\limits_{(r,\theta)}\!
\frac{1}{\sqrt{N}}\!\left| \sum _{n=-\frac{N-1}{2}}^{\frac{N-1}{2}}
\!\!\!\!\!\!e^{j\frac{2\pi n^2d^2}{c}\beta_{1,m}}
e^{-j\frac{2\pi n d}{c}\beta_{2,m}}
\right| \!\\&=\! \mathop{\mathrm{argmax}}\limits_{(r,\theta)}\frac{F\left(\frac{2\pi d^2}{c}\beta_{1,m},\frac{-2\pi d}{c}\beta_{2,m}\right)}{\sqrt{N}},
\end{aligned}
\end{split}
\end{equation}
where $\beta_{1,m} = f_m\frac{\cos^2{\theta}}{2{r}}-
\frac{
(W-\widetilde{f}_{m})f_0}{W} \frac{\cos^2\theta_{0}}{2r_0}-
\frac{
f_M\widetilde{f}_{m}}{W}\frac{\cos^2\theta_{c}}{2r_{c}}$
and $\beta_{2,m} = f_m\sin{\theta} -\frac{(W-\widetilde{f}_{m})f_0}{W}\sin\theta_0 - \frac{f_M\widetilde{f}_{m}}{W}\sin\theta_c$.
Since $F\left(\frac{2\pi d^2}{c}\beta_{1,m},\frac{-2\pi d}{c}\beta_{2,m}\right) \leq N$ and $F(0,0) = N$, considering the finite range of  near-field area, a feasible solution for $(\tilde{r}_m,\tilde{\theta}_m)$ is
\begin{align}
& \sin\tilde{{\theta}}_m=\frac{(W-\widetilde{f}_{m})f_0}{Wf_m}\sin\theta_{0}+\frac{(W+f_0)\widetilde{f}_{m}}{Wf_m}\sin\theta_{c},\label{1}\\
& \frac{1}{\tilde{r}_m}\!=\!\frac{1}{r_{0}}\!\frac{(W\!\!-\!\!\widetilde{f}_{m})f_0}{Wf_m}\frac{\cos^2{\theta}_0}{\cos^2\tilde{\theta}_m}\!+\!
\frac{1}{r_{c}}\!\frac{(W\!\!+\!\!f_0)\widetilde{f}_{m}}{Wf_m}\frac{\cos^2{\theta}_c}{\cos^2\tilde{\theta}_m}\label{2}.
\end{align}

\begin{figure}[!t]
\centering
\includegraphics[width=85mm]{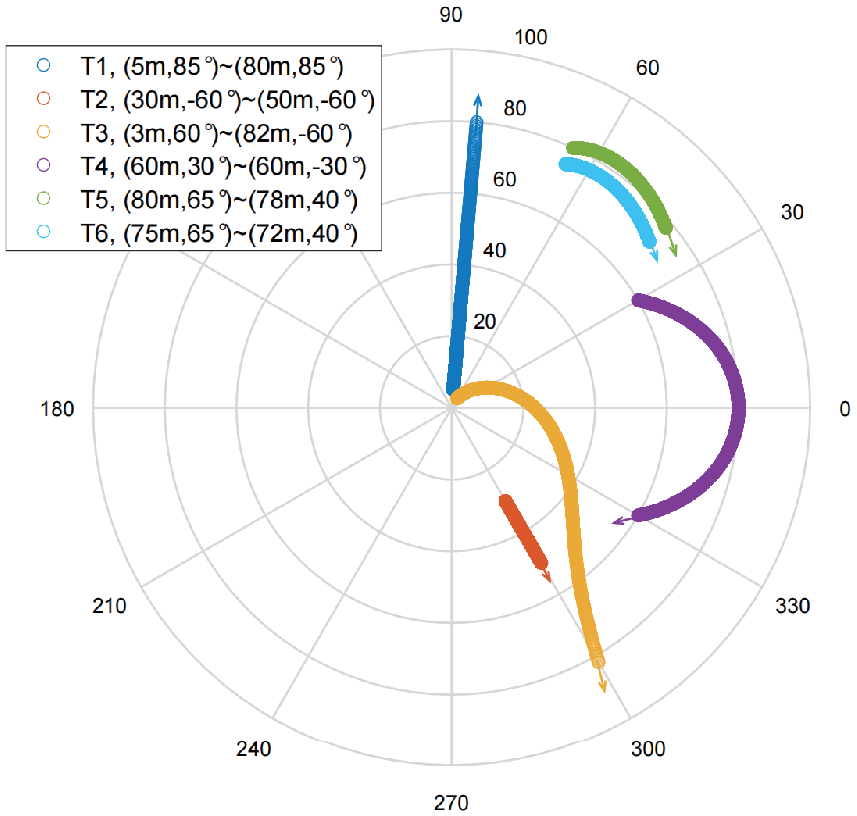}
\caption{TTDs-assisted  near-field controllable beam squint trajectories.}
\label{fig2:env}
\end{figure}

In summary, from the view of the BS,
we can autonomously control the start point and the end point of the squinted beams with the help of  TTDs,
such that different subcarriers will focus at different positions in the way we would expect. 
Some examples of different control schemes are shown in Fig.~5, where  $M=2048$ and $N=128$.
Trajectory T1 and T2 are two straight lines in the radial directions, where T1 squints from $(5m,85^\circ)$ to $(80m,85^\circ)$ while T2 squints from $(30m,-60^\circ)$ to $(50m,-60^\circ)$. 
If we already know the user's angle, then using the radius trajectories like T1 or T2 allows to quickly sense the user's distance.
Trajectory T3 squints from $(3m,60^\circ)$ to $(82m,-60^\circ)$, and the range of squint spans the entire expected sensing angle and distance regions.
Trajectory T4 squints from $(60m,30^\circ)$ to $(60m,-30^\circ)$ with
$t_{n,T4}\in[0.1889\mu s,0.2112\mu s]$ in a symmetrical form. This symmetry is more conducive to obtaining the user's angle.
Trajectories T5 and T6 show the controllable beam squint in a smaller range, 
which facilitates beam sweeping in a known narrow area.

\begin{figure}[!t]
\centering
\includegraphics[width=90mm]{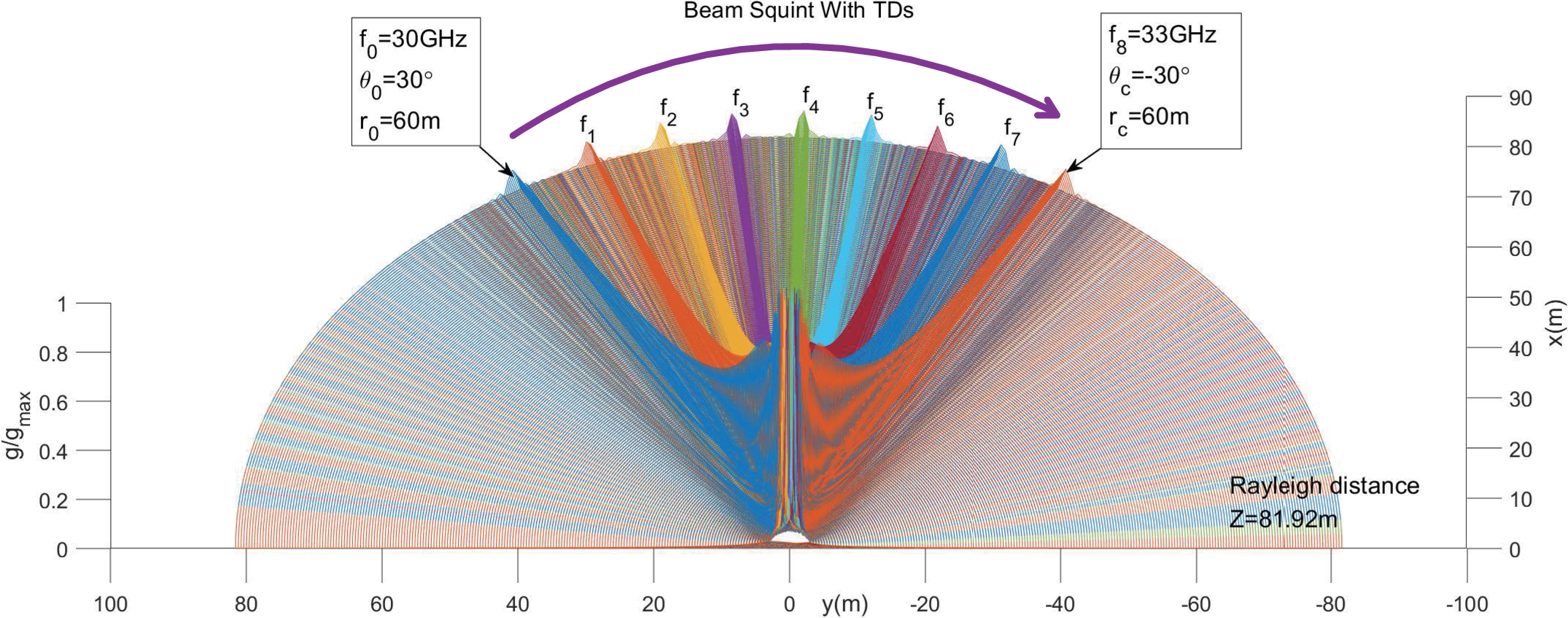}
\caption{Trajectory of TTDs-assisted beam squint. Where $f_0=$ 30 GHz, $W=$ 3 GHz, $M=8$, the lowest frequency carrier $f_0$ is set to focus on $(60m, 30^{\circ})$ and the highest frequency carrier $f_M$ is set to focus on $(60m, -30^{\circ})$.}
\label{fig_1}
\end{figure}

Let us now show how the trajectory T4 can obtain all users' angles.
We let  $M=8$ to clearly show the beamforming from different subcarriers.
The power of the received signal of each subcarrier at all positions can be calculated.
Fig. 6 shows the normalized received power of all discrete field points 
through the Electromagnetic distance $3.17m$ to Rayleigh distance $81.92m$ by distance step $0.4m$ and through angle range $90^\circ$ to $-90^\circ$ by angle step $-0.5^\circ$.
As the frequency gradually increases, the subcarrier squints from $30^\circ$ to $-30^\circ$, and different subcarriers have peaks in different directions. With this feature, we can determine the user's angle according to the subcarrier frequency of the maximum power fed back by the user, which will be densely effective when the number of subcarriers $M+1$ is large enough.

\section{Near-Field User Localization Scheme with Controllable Beam Squint}

\subsection{Baseline: Traditional  Beam Training Based Near-Field User Localization Scheme}

Beam training is an effective method for obtaining channel state information (CSI) in current communication systems. Through the beam training process between BS and users, the best beamforming vector can be selected from a set of pre-established near-field codebook. Due to the uniqueness of each codeword in the codebook, this process can also estimate the position of near-field users simultaneously.
Assume that the sensing range required by the BS is $\{(r,\theta)| r_{min}\leq r \leq r_{max}, \theta_{min}\leq \theta \leq \theta_{max}\}$,
and a traditional beam training based localization process can be divided into two stages: angle estimation and distance estimation.

During the angle estimation stage, the  angle sensing range can be uniformly divided into $I^{a}$ discrete angle values, forming $I^{a}$ discrete near-field angle units. The $i^a$-th  unit can be represented as $(r_a,\theta_{i^a})=
(r_a, \theta_{min}+(i^a-1)\Delta \theta)$, where $\Delta \theta=\frac{\theta_{max}-\theta_{min}}{I^a-1}$ and $i^a=1,2,...,I^a$.
Then  BS needs to implement $I^a$ times  beam sweeping based on PSs  for all  these  units, and 
 the beamforming vector  in the $i^a$-th beam sweeping is $\mathbf{w}(r_{a},\theta_{i^a})$. Assuming that  BS emits all $\mathbf{1}$ pilot signals,  the received  power of the $k$-th user on the $m$-th subcarrier can be denoted as $g_{k,m,i^a} = |\mathbf{h}^H_{k,m} \mathbf{w}(r_{a},\theta_{i^a})+n_{k,m,i^a}|$. Then the angle of the $k$-th user can be estimated as $\hat{\theta}_k^{Base} \!=\! \mathop{\mathrm{argmax}}\limits_{\theta_{i^a}}\sum_{m=0}^{M}g_{k,m,i^a}$.
In order to further estimate the distance of the $k$-th user, the BS divides the distance sensing range at $\hat{\theta}_k^{Base}$ direction into $I^d$ discrete distance values, forming $I^d$ near-field distance units.
The $i^{d,k}$-th unit is $(r_{i^d},\hat{\theta}_k^{Base})=(r_{min}+(i^{d,k}-1)\Delta r, \hat{\theta}_k^{Base})$, where $\Delta r = \frac{r_{max}-r_{min}}{I^d-1}$ and $i^{d,k}=1,2,...,I^d$.
Similar to the angle estimation process, the distance of the $k$-th user can be estimated as $\hat{r}_k^{Base} \!=\! \mathop{\mathrm{argmax}}\limits_{r_{i^d}}\sum_{m=0}^{M}|\mathbf{h}^H_{k,m} \mathbf{w}(r_{i^d},\hat{\theta}_k^{Base})+n_{k,m,i^d}|$.

For ease of expression, this beam training based scheme will be abbreviated as \emph{TBT-Baseline} in this paper. 
Due to the fact that this traditional scheme can only concentrate the beam on one near-field position in  one time beam sweeping, the TBT-Baseline scheme requires a total of $I=I^a+K\times I^d$ times beam sweeping to obtain the positions of all $K$ users.

\subsection{Proposed: A Low Complexity  User Localization Scheme Based on Controllable Beam Squint}

With the help of TTDs, the BS can adjust the values  of the PSs as $\phi_{n}=\frac{f_0r_{0,n}}{c}$ and the values of the TTDs as $t_{n}=\frac{f_M}{Wc}r_{c,n}-\frac{\phi_{n}}{W}$ to make the
beamforming of the subcarriers from the $0$-th subcarrier to the $M$-th subcarrier
squint from the start position
 $(r_0,\theta_0)$ to the
 end position $(r_c,\theta_c)$.
Assuming the BS emits all $\mathbf{1}$ pilot signals,
under ideal noiseless conditions,
  the complex received signal of the $k$-th user on the $m$-th subcarrier now
 can be represented as
\begin{equation}
\begin{split}
\begin{aligned}
\label{deqn_ex1a}
\tilde{y}_{k,m}&= \mathbf{h}^H(r_k,\theta_k,f_m) \cdot \tilde{\mathbf{w}}\\&=
\frac{\alpha_{k,m}}{\sqrt{N}}\sum_{n=-\frac{N-1}{2}}^{\frac{N-1}{2}}
e^{j2\pi f_m\frac{r_{k,n}}{c}}
e^{-j2\pi \phi_n}e^{-j2\pi \tilde{f}_m t_n}.
\end{aligned}
\end{split}
\end{equation}
For the convenience of  analysis, we multiply $\tilde{y}_{k,m}$ by $\frac{4\pi\sqrt{N}}{\lambda _m}$ 
to eliminate the influence of different subcarriers in $\alpha_{k,m}$
and  obtain the received  signal after being processed as
\begin{equation}
\begin{split}
\begin{aligned}
\label{deqn_ex1a}
\!\!\!\!\breve{y}_{k,m}\!\! &= \frac{4\pi\sqrt{N}}{\lambda _m} \tilde{y}_{k,m}
\!\\&=\!\!\frac{1}{r_k}\!\!\sum_{n=-\frac{N-1}{2}}^{\frac{N-1}{2}}\!\!\!\!\!\!
e^{j\!\frac{2\pi f_m}{c}r_{k,n}}
e^{-j\!\frac{2\pi (W\!-\!\tilde{f}_m)f_0}{Wc}r_{0,n}}
e^{-j\!\frac{2\pi \tilde{f}_m f_M}{Wc}r_{c,n}}\\
&=\frac{1}{r_k}e^{j\xi_{0,k,m}}\sum_{n=-\frac{N-1}{2}}^{\frac{N-1}{2}}
e^{jn\xi_{1,k,m}}
e^{jn^2\xi_{2,k,m}},
\end{aligned}
\end{split}
\end{equation}
where $\xi_{0,k,m}=\frac{2\pi}{Wc}[Wf_mr_k-(W-\tilde{f}_m)f_0r_0 - \tilde{f}_mf_Mr_c]$,
$\xi_{1,k,m} = -\frac{2\pi d}{Wc}[Wf_m\sin\theta_k-(W-\tilde{f}_m)f_0\sin\theta_0 - \tilde{f}_mf_M\sin\theta_c]$ and $\xi_{2,k,m} = \frac{2\pi d^2}{Wc}[Wf_m \frac{\cos^2\theta_k}{2r_k}-(W-\tilde{f}_m)f_0 \frac{\cos^2\theta_0}{2r_0} - \tilde{f}_mf_M \frac{\cos^2\theta_c}{2r_c}]$.

Assume that
the sensing range required by the BS is still  $\{(r,\theta)| r_{min}\leq r \leq r_{max}, \theta_{min}\leq \theta \leq \theta_{max}\}$.
The proposed low complexity localization scheme includes two stages: angle sensing and distance sensing. 
In the angle sensing stage, assume that
 $r_{mid1}$, $r_{mid2}\in[r_{min},r_{max}]$ are two appropriate specific distance values.
We target to use one time beam sweeping to get all $K$ users' angle estimation results $\hat{\theta}_k^{Pro1}$, where $k=1,2,...,K$.
The specific steps are as follows:

Firstly, by adjusting the values of PSs and TTDs,  the beamforming of the $0$-th and $M$-th subcarriers can be controlled to focus on  positions $(r_{mid1},\theta_{max})$ and $(r_{mid2},\theta_{min})$, respectively.
Secondly, the
 BS simultaneously transmits over $M+1$ subcarriers, whose
 angles squint from $\theta_{max}$ to $\theta_{min}$
and  distances squint from $r_{mid1}$ to $r_{mid2}$, 
covering the whole space in the form of a curve, in which
the trajectory is similar to trajectory T4 in Fig. 5.
Thirdly, all users will receive $M+1$ subcarriers, and  
 the  power of the $m$-th subcarrier received by the $k$-th user is
$\breve{g}_{k,m}^{angle} = |\breve{y}_{k,m}^{angle}|=\frac{1}{r_k}|\sum_{n=-\frac{N-1}{2}}^{\frac{N-1}{2}}
e^{jn\xi_{1,k,m}^{angle}}
e^{jn^2\xi_{2,k,m}^{angle}}|$.
Substituting $r_0=r_{mid1}$, $\theta_0=\theta_{max}$, $r_c=r_{mid2}$ and $\theta_c=\theta_{min}$ into $\xi_{1,k,m}$ and $\xi_{2,k,m}$,  we can obtain 
$\xi_{1,k,m}^{angle} = -\frac{2\pi d}{Wc}[Wf_m\sin\theta_k-(W-\tilde{f}_m)f_0\sin\theta_{max} - \tilde{f}_mf_M\sin\theta_{min}]$ and $\xi_{2,k,m}^{angle} = \frac{2\pi d^2}{Wc}[Wf_m \frac{\cos^2\theta_k}{2r_k}-(W-\tilde{f}_m)f_0 \frac{\cos^2\theta_{max}}{2r_{mid1}} - \tilde{f}_mf_M \frac{\cos^2\theta_{min}}{2r_{mid2}}]$.
In order to simplify the second-order calculation of the near-field, we have to select appropriate $r_{mid1}$ and ${r_{mid2}}$, such that there is $\xi_{2,k,m}^{angle}\approx p_{k,m}^{angle}\cdot 2\pi$ within the sensing angle range,
 where $p_{k,m}^{angle}$ is an integer introduced for approximate phase alignment.
Then $\breve{g}_{k,m}^{angle}$ can be reduced to
\begin{equation}
\begin{split}
\begin{aligned}
\label{deqn_ex1a}
\breve{g}_{k,m}^{angle}\!
=\frac{1}{r_k}\left|\sum _{n=-\frac{N-1}{2}}^{\frac{N-1}{2}}\! e^{jn\xi_{1,k,m}^{angle}}\right|
=\frac{1}{r_k}\left|\frac{\sin(\frac{\xi_{1,k,m}^{angle}}{2}N)}{\sin(\frac{\xi_{1,k,m}^{angle}}{2})}\right|.
\end{aligned}
\end{split}
\end{equation}

\begin{figure}[!t]
\centering
\includegraphics[width=90mm]{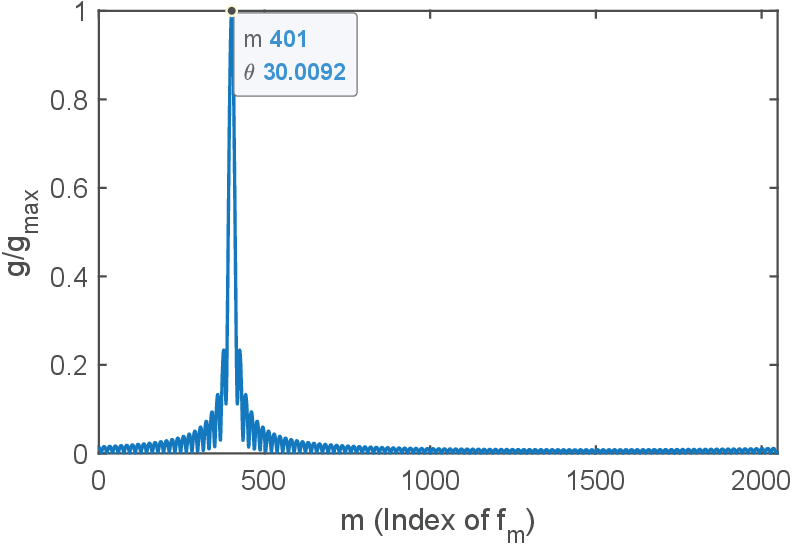}
\caption{Normalized power spectrum of received signal of the user at $(30m,30^\circ)$ in angle sensing stage of the proposed low complexity localization scheme. $M=2048$ and $N=128$.}
\label{fig_1}
\end{figure}

Since $\xi_{1,k,m}^{angle}$ is a linear function of $f_m$
and the product of its two endpoints is $\xi_{1,k,0}^{angle}\cdot \xi_{1,k,M}^{angle}=\frac{4\pi^2 d^2 f_0 f_M}{c^2}(\sin\theta_k - \sin\theta_{max})(\sin\theta_k-\sin\theta_{min})< 0$, when $M+1$ is large enough, there must be an approximate $f_{m_k}^{angle}$ to make
$\xi_{1,k,m_k}^{angle} = 0$. This means that there will be a peak value $\breve{g}_{k,m_k}^{angle}$ in the revised power spectrum of the $k$-th user $\mathbf{\breve{g}}_k^{angle} = [\breve{g}_{k,0}^{angle},\breve{g}_{k,1}^{angle},...,\breve{g}_{k,M}^{angle}]^T$.
We call this subcarrier as the \emph{maximum power subcarrier} corresponding to the $k$-th user in angle sensing stage, and record its frequency as $f_{k,d_\theta}^{Pro1}=f_{m_k}^{angle}$.
Then 
the $k$-th user  can feed back  $f_{k,d_\theta}^{Pro1}$ to the BS,
and the BS can calculate the angle estimation of the $k$-th user as $\hat{\theta}_k^{Pro1}$
 by substituting 
$\tilde{\theta}_m=\hat{\theta}_k^{Pro1}$, $f_m=f_{k,d_\theta}^{Pro1}$ , $\widetilde{f}_m=f_{k,d_\theta}^{Pro1}-f_0$, $\theta_0 = \theta_{max}$ and $\theta_c = \theta_{min}$ into (19). By doing so, we can obtain the angle estimation results for each user.

Although we make some approximations when calculating (23),
Fig. 7 shows the simulation results  using formula (22) without approximation.
When the user is at $(30m,30^\circ)$,
we can get that the $401$-th subcarrier is the maximum power subcarrier corresponding to this user in angle sensing stage. Then the BS can 
 calculate the user's angle as $\hat{\theta}_{example}^{Pro1}=30.0092^\circ$ from $f_{example,d_\theta}^{Pro1}=f_{401}$.

It should be noted that if more than two users are in the same angle direction, 
then their angle estimates will be the same. 
Hence the angle sensing stage will get $\check{K}$ different angle estimates,
with $\check{K}\leq K$ and $\check{K}\leq M+1$. Let us mark these $\check{K}$ different angles as 
$\check{\theta}_1,\check{\theta}_2,...,\check{\theta}_{\check{K}}$.
Then the BS needs to perform $\check{K}$ times of TTDs-assisted
controllable 
 beam squint to complete the distance sensing for all users.
Assuming  that $\hat{\theta}_k^{Pro1} = \check{\theta}_{\check{k}}$,
we next sense the $k$-th  user's distance $\hat{r}_k^{Pro1}$ in the $\check{k}$-th 
beam sweeping with the following steps:

Firstly, by adjusting the values of PSs and TTDs,  the beamforming of the $0$-th and $M$-th subcarriers can be controlled to focus  on  $(r_{min},\hat{\theta}_k^{Pro1})$ and $(r_{max},\hat{\theta}_k^{Pro1})$, respectively.
Secondly, the BS simultaneously transmits over $M+1$ subcarriers, whose beamforming
 angles are $\hat{\theta}_k^{Pro1}$,
but the distances squint from $r_{min}$ to $r_{max}$, 
covering the whole space in the form of a straight segment,
which is similar to trajectory T1 in Fig. 5. 
Thirdly, the $k$-th user will receive $M+1$ subcarriers, and  
 the  power of the $m$-th subcarrier received by the $k$-th user is
$\breve{g}_{k,m}^{distance} = |\breve{y}_{k,m}^{distance}|=\frac{1}{r_k}|\sum_{n=-\frac{N-1}{2}}^{\frac{N-1}{2}}
e^{jn\xi_{1,k,m}^{distance}}
e^{jn^2\xi_{2,k,m}^{distance}}|$.
Substituting $\theta_k = \hat{\theta}_k^{Pro1}$, $r_0=r_{min}$, $\theta_0=\hat{\theta}_k^{Pro1}$, $r_c=r_{max}$ and $\theta_c=\hat{\theta}_k^{Pro1}$ into $\xi_{1,k,m}$ and $\xi_{2,k,m}$,  we can obtain 
$\xi_{1,k,m}^{distance} = 0$ and $\xi_{2,k,m}^{distance} = \frac{\pi d^2\cos^2 \hat{\theta}_k^{Pro1}}{Wc}[Wf_m \frac{1}{r_k}-(W-\tilde{f}_m)f_0 \frac{1}{r_{min}} - \tilde{f}_mf_M \frac{1}{r_{max}}]$.
Then $\breve{g}_{k,m}^{distance}$ can be reduced to
\begin{equation}
\begin{split}
\begin{aligned}
\label{deqn_ex1a}
\breve{g}_{k,m}^{distance}\!
=\frac{1}{r_k}\left|\sum _{n=-\frac{N-1}{2}}^{\frac{N-1}{2}}\! e^{jn^2\xi_{2,k,m}^{distance}}\right|.
\end{aligned}
\end{split}
\end{equation}

Since $\xi_{2,k,m}^{distance}$ is a linear function of $f_m$
and the product of its two endpoints is $\xi_{2,k,0}^{distance}\cdot \xi_{2,k,M}^{distance}=\frac{\pi^2 d^4 \cos^4 \hat{\theta}_k^{Pro1} f_0 f_M}{c^2}(\frac{1}{r_k} - \frac{1}{r_{min}})(\frac{1}{r_k} - \frac{1}{r_{max}})< 0$, when $M+1$ is large enough, there must be an approximate $f_{m_k}^{distance}$ to make
$\xi_{2,k,m_k}^{distance} = 0$. This means that there will be a peak value $\breve{g}_{k,m_k}^{distance}$ in the revised power spectrum of the $k$-th user $\mathbf{\breve{g}}_k^{distance} = [\breve{g}_{k,0}^{distance},\breve{g}_{k,1}^{distance},...,\breve{g}_{k,M}^{distance}]^T$.
We record the frequency of the maximum power subcarrier corresponding to the $k$-th user in the distance sensing stage as $f_{k,d_r}^{Pro1} = f_{m_k}^{distance}$.
After the $k$-th user provides $f_{k,d_r}^{Pro1}$ feedback to the BS,
the BS can calculate the distance estimation of the $k$-th user as $\hat{r}_k^{Pro1}$
 by substituting 
$\tilde{r}_m=\hat{r}_k^{Pro1}$, $f_m=f_{k,d_r}^{Pro1}$ , $\widetilde{f}_m=f_{k,d_r}^{Pro1}-f_0$, $r_0 = r_{min}$ , $r_c = r_{max}$, $\tilde{\theta}_m =\hat{\theta}_k^{Pro1}$ and $\theta_0 = \theta_c =  \hat{\theta}_k^{Pro1}$ into (20). By doing so, we can obtain the distance estimation results for each user.

\begin{figure}[!t]
\centering
\includegraphics[width=90mm]{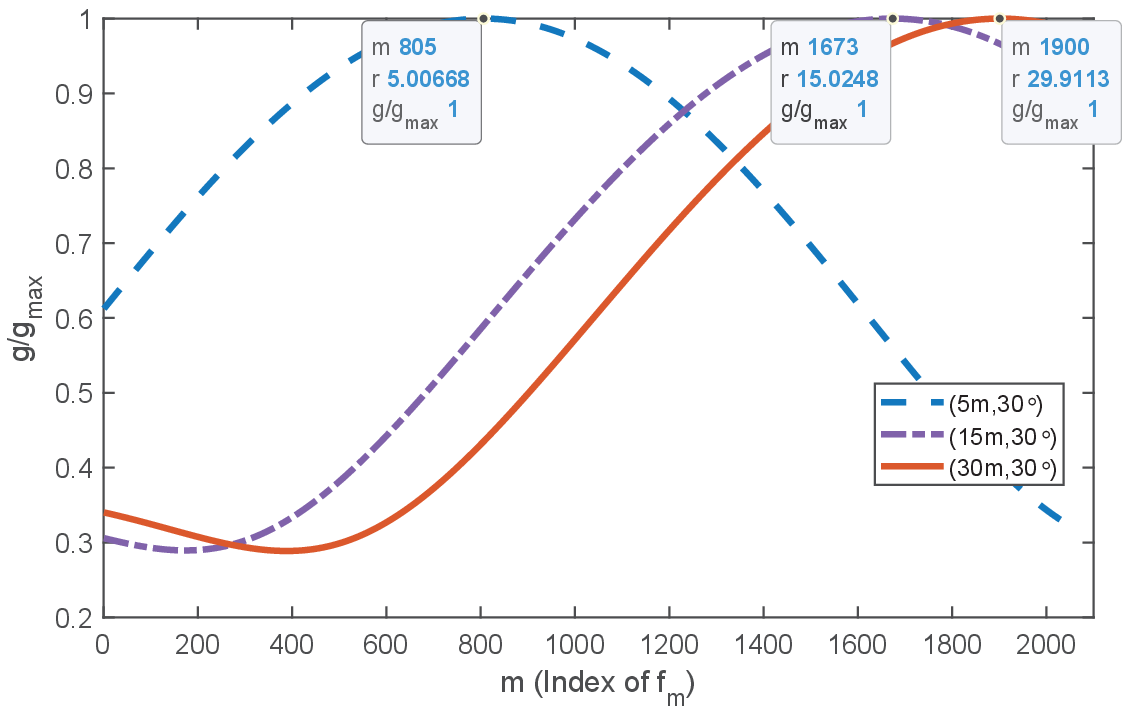}
\caption{Normalized power spectrums of received signals of the users at $(5m,30^\circ)$, $(15m,30^\circ)$ and $(30m,30^\circ)$ position in distance sensing stage of the proposed low complexity localization scheme. $M=2048$ and $N=128$.}
\label{fig_1}
\end{figure}

An example of  distance sensing is shown in Fig.~8.
For the  user at $(30m,30^\circ)$, we can get that the $1900$-th subcarrier has the maximum power,
and then calculate the user's distance as $\hat{r}_{example}^{Pro1}=29.9113m$ from $f_{example,d_r}^{Pro1} = f_{1900}$.
Furthermore, the normalized power spectrums of another two users in  direction $30^\circ$ are also given in Fig. 8.
We can see that the users at different distances receive the maximum power at different subcarriers, which allows BS to determine the user's distance.
However,
note that the power change of different subcarriers near the maximum power is flat, which is very different from the angle sensing as shown in Fig.~7.
This means that the distance sensing in the proposed low complexity localization scheme is susceptible to the noise interference in environments with low signal-to-noise (SNR) ratio.

\subsection{Proposed: A High Accuracy User Localization Scheme Based on Controllable Beam Squint}

We further propose a high-accuracy localization scheme based on controllable beam squint. Specifically, the BS needs to implement $P$ times TTDs-assisted beamforming to sense the positions of all users.
In the $p$-th beam sweeping, by adjusting the values of PSs and TTDs, we can focus the beamforming of the $0$-th and the $M$-th subcarriers on positions $(r_{mid1},\theta_{max,p})$ and $(r_{mid2},\theta_{min,p})$, respectively.
In addition, we request that $\min\{\theta_{max,1},\theta_{max,2},...,\theta_{max,P}\}\geq \theta_{max}$ , $\max\{\theta_{min,1},\theta_{min,2},...,\theta_{min,P}\}\leq \theta_{min}$ , $\theta_{max,i}\neq \theta_{max,j}$ and $\theta_{min,i}\neq \theta_{min,j}$ for $i \neq j$, such  that every time beam sweeping can cover the required angle sensing range, but the range of each beam sweeping is slightly different.

Based on formula (22), we can easily represent the  received signal  after being processed on the $m$-th subcarrier of the $k$-th user in the $p$-th beam sweeping as
\begin{equation}
\begin{split}
\begin{aligned}
\label{deqn_ex1a}
\breve{y}_{k,m,p}^{Pro2}
=\frac{1}{r_k}e^{j\xi_{0,k,m,p}^{Pro2}}\sum_{n=-\frac{N-1}{2}}^{\frac{N-1}{2}}
e^{jn\xi_{1,k,m,p}^{Pro2}}
e^{jn^2\xi_{2,k,m,p}^{Pro2}},
\end{aligned}
\end{split}
\end{equation}
where $\xi_{0,k,m,p}^{Pro2}=\frac{2\pi}{Wc}[Wf_mr_k-(W-\tilde{f}_m)f_0r_{mid1} - \tilde{f}_mf_Mr_{mid2}]$,
$\xi_{1,k,m,p}^{Pro2} = -\frac{2\pi d}{Wc}[Wf_m\sin\theta_k-(W-\tilde{f}_m)f_0\sin\theta_{max,p} - \tilde{f}_mf_M\sin\theta_{min,p}]$ and $\xi_{2,k,m,p}^{Pro2} = \frac{2\pi d^2}{Wc}[Wf_m \frac{\cos^2\theta_k}{2r_k}-(W-\tilde{f}_m)f_0 \frac{\cos^2\theta_{max,p}}{2r_{mid1}} - \tilde{f}_mf_M \frac{\cos^2\theta_{min,p}}{2r_{mid2}}]$.
On the one hand, according to the analysis of the angle sensing stage in the proposed low complexity localization scheme, there must be one peak value in the  received signal power spectrum of the $k$-th user in the $p$-th beam sweeping, whose corresponding maximum power subcarrier frequency can be recorded as $f_{k,d_\theta,p}^{Pro2}$, and the user's angle will be estimated as $\hat{\theta}_{k,p}^{Pro1}$.
Besides, we
mark the index of this subcarrier as $m'_{k,p}$, and
the theoretical phase of the $k$-th user's maximum power subcarrier in the $p$-th beam sweeping can be expressed as $\phi_{k,p}^{Pro2} = \arg\{\breve{y}_{k,m'_{k,p},p}^{Pro2}\}$. On the other hand, users can detect the phase on each subcarrier of their received signal through the built-in channel detector. Record the phase measurement value of the $k$-th user's maximum power subcarrier in the $p$-th beam sweeping as $\varphi_{k,p}^{Pro2}$.
Therefore, the $k$-th user can feedback $f_{k,d_\theta,p}^{Pro2}$ and $\varphi_{k,p}^{Pro2}$ to the BS.

After  $P$ times beam sweeping, the BS will receive the information about the $k$-th user as $\{f_{k,d_\theta,1}^{Pro2},f_{k,d_\theta,2}^{Pro2},...,f_{k,d_\theta,P}^{Pro2}\}$ and $\{\varphi_{k,1}^{Pro2},\varphi_{k,2}^{Pro2},...,\varphi_{k,P}^{Pro2}\}$, and 
 BS can calculate the user's angle estimation result in each time beam sweeping as
$\{\hat{\theta}_{k,1}^{Pro1},\hat{\theta}_{k,2}^{Pro1},...,\hat{\theta}_{k,P}^{Pro1}\}$.
Then a higher accuracy angle estimation result for the $k$-th user can be obtained
by taking the average value as
\begin{equation}
\begin{split}
\begin{aligned}
\label{deqn_ex1a}
\hat{\theta}_{k}^{Pro2} = \frac{1}{P}\sum_{p=1}^P \hat{\theta}_{k,p}^{Pro1}.
\end{aligned}
\end{split}
\end{equation}

In order to further improve the accuracy of distance estimation,
 we introduce the multiple carrier phase difference based ranging method.
Due to the different phase changes caused by subcarriers of different frequencies transmitting the same distance, the transmission distance can be estimated based on the phase differences between multiple carriers\cite{2022arXiv220602996S,DBLP23}.
Due to the slightly different angle squint range set for each time beam sweeping, the frequency of the $k$-th user's maximum power subcarrier will vary during each beam sweeping. Hence the distance estimation result for the $k$-th user can be obtained through  one-dimensional distance search as
\begin{equation}
\begin{split}
\begin{aligned}
\label{deqn_ex1a}
\hat{r}_{k}^{Pro2} = \mathop{\mathrm{argmax}}\limits_{r} L(r)
= \mathop{\mathrm{argmax}}\limits_{r}
\left|\sum_{p=1}^P
e^{j[\varphi_{k,p}^{Pro2} - \phi_{k,p,r}^{Pro2}]}
\right|,
\end{aligned}
\end{split}
\end{equation}
with 
\begin{equation}
\begin{split}
\begin{aligned}
\label{deqn_ex1a}
&\!\!\!\!\!\phi_{k,p,r}^{Pro2} = \arg\{\breve{y}_{k,m'_{k,p},p,r}^{Pro2}\}\\\!\!\!&\!\!\!=\!
\arg\!\left\{\!\!e^{j\xi_{0,k,m'_{k,p},p,r}^{Pro2}}\!\!\!\!\!\!\!\!\sum_{n=-\frac{N-1}{2}}^{\frac{N-1}{2}}\!\!\!\!\!\!
e^{jn\xi_{1,k,m'_{k,p},p,r}^{Pro2}}\!
e^{jn^2\xi_{2,k,m'_{k,p},p,r}^{Pro2}}\!\!\right\},
\end{aligned}
\end{split}
\end{equation}
where $\xi_{1,k,m'_{k,p},p,r}^{Pro2} = -\frac{2\pi d}{Wc}[Wf_{k,d_\theta,p}^{Pro2}\sin\hat{\theta}_{k,p}^{Pro1}-(W-\tilde{f}_{k,d_\theta,p}^{Pro2})f_0\sin\theta_{max,p} - 
\tilde{f}_{k,d_\theta,p}^{Pro2}f_M\sin\theta_{min,p}] = 0$ (since $\hat{\theta}_{k,p}^{Pro1}$ is calculated by $f_{k,d_\theta,p}^{Pro2}$ and (19)),
$\xi_{0,k,m'_{k,p},p,r}^{Pro2} = \frac{2\pi}{Wc}[Wf_{k,d_\theta,p}^{Pro2}r-(W-\tilde{f}_{k,d_\theta,p}^{Pro2})f_0r_{mid1} - \tilde{f}_{k,d_\theta,p}^{Pro2}f_Mr_{mid2}] $, 
$\xi_{2,k,m'_{k,p},p,r}^{Pro2} = \frac{2\pi d^2}{Wc}[Wf_{k,d_\theta,p}^{Pro2} \frac{\cos^2\hat{\theta}_{k,p}^{Pro1}}{2r}-(W-\tilde{f}_{k,d_\theta,p}^{Pro2})f_0 \frac{\cos^2\theta_{max,p}}{2r_{mid1}} - \tilde{f}_{k,d_\theta,p}^{Pro2}f_M \frac{\cos^2\theta_{min,p}}{2r_{mid2}}]$.

\begin{figure*}[!t]
\centering
\subfloat[]{\includegraphics[width=80mm]{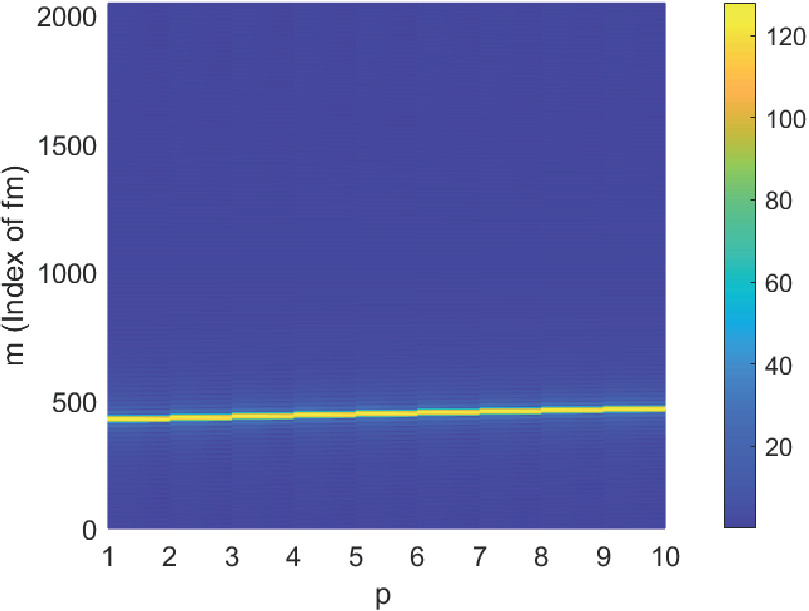}%
\label{fig_first_case}}
\hfil
\subfloat[]{\includegraphics[width=77mm]{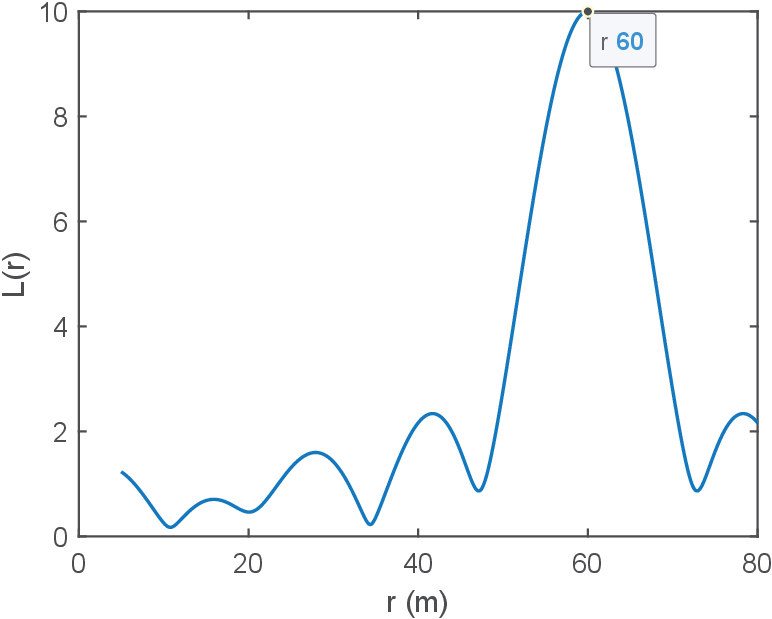}%
\label{fig_second_case}}
\caption{An example of the proposed high-accuracy user localization scheme, in which the user is at $(60m, 30^\circ)$, $M=2048$, $N=128$ and $P=10$. (a) The power spectrum matrix composed of the power spectrum received by the user during each beam sweeping.
(b) Proposed one-dimensional distance search curve for user distance sensing.}
\label{fig_sim}
\end{figure*}

Equation (27) computes the difference between the measured phases and the theoretical phases over distance in  peak power subcarriers.  Essentially, the optimal distance estimation $\hat{r}_{k}^{Pro2}$ leads to the closest match between these two sets of phase values. Therefore, we can estimate the distance of the user by plotting  $r$-$L(r)$ curve and searching for the peak.

An example of the proposed high-accuracy user localization scheme is shown in Fig.~9, where the user is located at $(60m, 30^\circ)$.
Fig.~9 (a) shows the power spectrum matrix composed of the received signal power spectrum of the user in each time beam sweeping.
It can be seen that the index of the maximum power subcarrier in the user's received signal is slightly different during each beam sweeping, and the maximum power subcarrier approximately forms a clear high power line in the entire spectrum matrix.
Using the phase measurement values of the maximum power subcarriers with different frequencies, Fig.~9 (b) shows the one-dimensional distance search curve based on  (27), in which an obvious peak corresponding to the search distance $60m$ can be found. Hence we can estimate the distance of the user as $60m$.

\subsection{Proposed: Joint Localization Scheme Based on Double BSs and Controllable Beam Squint}

\begin{figure}
\centering
\includegraphics[width=80mm]{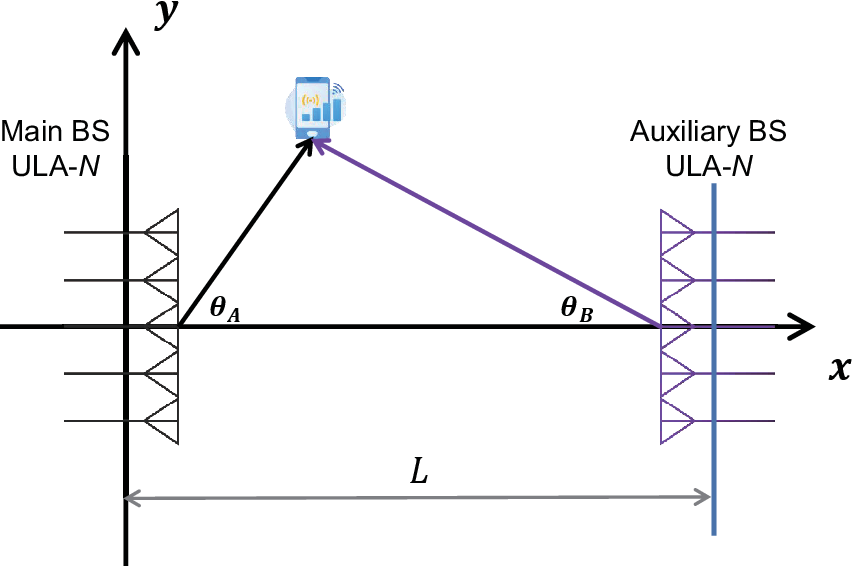}
\caption{Double BSs system.}
\label{fig1:env}
\end{figure}

The proposed high-accuracy localization scheme requires the one-dimensional distance search for each user during distance sensing, which may result in significant computational complexity.
In order to reduce the computational complexity while also reducing
 the cost of beam sweeping times,
 we consider employing two BSs for  user localization   in ISAC system.
The structure and principle of the double BSs localization system are shown in Fig. 10.
It is assumed that both the two BSs are equipped with the ULA of $N$ antennas, and the setting of the main BS is completely the same as the system model in Section \RNum{2}.
The newly added auxiliary BS is parallel to the main BS, and its array center is located at $(L,0)$. Other parameters of the auxiliary BS are identical to those of the main BS.

For the $k$-th user, 
both the two BSs can use the angle sensing stage proposed in the 
low complexity   localization scheme
 to obtain the user's angles with respect to each BS, which are denoted as $\theta_{A,k}$ and $\theta_{B,k}$ here. 
Then the angle and distance estimation results of the $k$-th user can be calculated according to the  geometric relationship and trigonometric operation as
\begin{align}
& \hat{\theta}_{k}^{Pro3}=\theta_{A,k},\label{1}\\
& \hat{r}_{k}^{Pro3}=\frac{L}{\cos\theta_{A,k}+\frac{\sin\theta_{A,k}}{\tan\theta_{B,k}}}\label{2}.
\end{align}

\subsection{Resource Overhead Analysis}

We  define the abbreviations of the four localization schemes involved in this paper as shown in Table~$\rm \uppercase\expandafter{\romannumeral1}$.
In addition, the resource overhead of the four schemes are summarized in Table~$\rm \uppercase\expandafter{\romannumeral2}$, where $P_R$ is the number of  search grids for  distance search in CBS-High                                                                                    Scheme.

\begin{table}[h]
\centering
\caption{List of Abbreviated Names for the Schemes}
\begin{tabular}{|>{\hspace{0pt}}m{0.62\linewidth}|>{\hspace{0pt}}m{0.185\linewidth}|} 
\hline
\multicolumn{1}{|>{\centering\hspace{0pt}}m{0.62\linewidth}|}{Scheme}                 & \multicolumn{1}{>{\centering\arraybackslash\hspace{0pt}}m{0.185\linewidth}|}{Abbreviation}  \\ 
\hline
Baseline: Traditional  Beam Training Based Near-Field User Localization Scheme         & TBT-Baseline                                                                                 \\ 
\hline
Proposed: A Low Complexity  User Localization Scheme Based on Controllable Beam Squint & CBS-Low                                                                                     \\ 
\hline
Proposed: A High Accuracy User Localization Scheme Based on Controllable Beam Squint   & CBS-High                                                                                    \\ 
\hline
Proposed: Joint Localization Scheme Based on Double BSs and Controllable Beam Squint   & CBS-2BS                                                                                     \\
\hline
\end{tabular}
\end{table}

\begin{table}
\centering
\caption{Overhead Analysis}
\begin{tabular}{|>{\centering\hspace{0pt}}m{0.11\linewidth}|>{\centering\hspace{0pt}}m{0.214\linewidth}|>{\centering\hspace{0pt}}m{0.1815\linewidth}|>{\centering\hspace{0pt}}m{0.15\linewidth}|>{\centering\arraybackslash\hspace{0pt}}m{0.11\linewidth}|} 
\hline
Scheme       & Beam sweeping times\par{}for all users localization & Number of feedback\par{}values per user & Computing times\par{}per user & Number of BSs  \\ 
\hline
TBT-Baseline & $I=I^a+K\times I^d$          & $2$                                       & $2$                             & $1$              \\ 
\hline
CBS-Low      & $K+1$                                                 & $2$                                       & $2$                             & $1$              \\ 
\hline
CBS-High     & $P$                                                   & $2P$                                      & $P_R+P+1$                      & $1$              \\ 
\hline
CBS-2BS      & $2$                                                   & $2$                                       & $3$                             & $2$              \\
\hline
\end{tabular}
\end{table}

In terms of beam sweeping cost, the localization schemes based on controllable beam squint can scan $M+1$ near-field positions in one time beam sweeping, while the TBT-Baseline Scheme can only scan one near-field position  in one time beam sweeping. 
In order to ensure that the spatial scanning interval between the TBT-Baseline Scheme and the CBS Schemes is  basically consistent, we set  $I^a=M+1$ and $I^d=M+1$, respectively. Then the  TBT-Baseline Scheme requires a total of $I=(K+1)(M+1)$ times beam sweeping for all users localization.
In specific practice, there is $I \gg K+1$, $I \gg P$ and $I \gg 2$. (More detailed numerical results can be found in the simulation section.)
This means that the proposed beam squint based CBS-Low Scheme, CBS-High Scheme and CBS-2BS Scheme  greatly reduce the overhead of beam sweeping times for sensing, which mainly due to the fact that the controllable beam squint can  cover multiple near-field positions within only one OFDM symbol.
Besides, the CBS-High Scheme improves the localization accuracy  at the cost of computational complexity.
The proposed CBS-2BS  Scheme  realize high-precision localization  with low computational complexity and minimal beam sweeping times  at the cost of adding an additional auxiliary BS.

In terms of user feedback cost, the localization schemes mentioned in this paper belong to cooperative user sensing schemes based on  downlink detection and  uplink feedback, both of which require user feedback. 
The TBT-Baseline Scheme requires each user to provide a total of $2$ maximum power indices for feedback during the angle and distance estimation process. The CBS-Low Scheme requires each  user to provide 
$2$ frequency values of the maximum power subcarriers for feedback during the angle and distance sensing stages.
The CBS-2BS Scheme requires each user to provide feedback with a total of $2$ frequency values of the maximum power subcarriers for double BSs angles sensing. This means that the feedback costs of these three schemes are consistent.
In addition,   each user needs to provide feedback to the BS with $P$ frequencies value and $P$ phase values in CBS-High   Scheme. However, we can see in the simulation that $P$ is usually not set very large, and its typical value is below $10$. Therefore, we believe that compared to the significantly reduced sensing beam sweeping times overhead and higher localization accuracy brought by  CBS-High   Scheme, the introduction of $2P-2$ additional user feedback values can be accepted.

\section{Simulation Results }

In simulations,
 root mean square error (RMSE) is employed to characterize the sensing performance.
The RMSE of angle sensing, distance sensing and 2D localization can be defined as
${\rm RMSE_\theta}=\sqrt{\frac{\sum _{t=1}^{T}(\hat{\theta}_{sp,i}-\theta_{sp})^2}{T}}$, ${\rm RMSE_r}=\sqrt{\frac{\sum _{t=1}^{T}(\hat{r}_{sp,i}-r_{sp})^2}{T}}$ and ${\rm RMSE_{2D}}=\sqrt{\frac{\sum _{t=1}^{T}[(\hat{x}_{sp,t}-x_{sp})^2+(\hat{y}_{sp,t}-y_{sp})^2]}{T}}$,
where $T$ is the number of repeated experiments, $(r_{sp},\theta_{sp})$ and $(x_{sp},y_{sp})$ are the user's real position, and $(\hat{r}_{sp,t},\hat{\theta}_{sp,t})$ and $(\hat{x}_{sp,t},\hat{y}_{sp,t})$ are the estimations of the user's position.
Assume that the received signal on the $m$-th subcarrier  is
$y_m^{noise} = y_m^{no} + n_m^{noise}$, where $y_m^{no}$ is a signal without  noise, $n_m^{noise}$ is the zero-mean additive complex Gaussian white noise, and $y_m^{noise}$ is the signal after adding noise.
Then the signal-to-noise  ratio (SNR)  can be defined as
${\rm SNR}=\frac{\mathbb E\{|y_m^{no}|^2\}}{\mathbb E\{|n_m^{noise}|^2\}}$.

\begin{figure*}[!t]
\centering
\subfloat[]{\includegraphics[width=80mm]{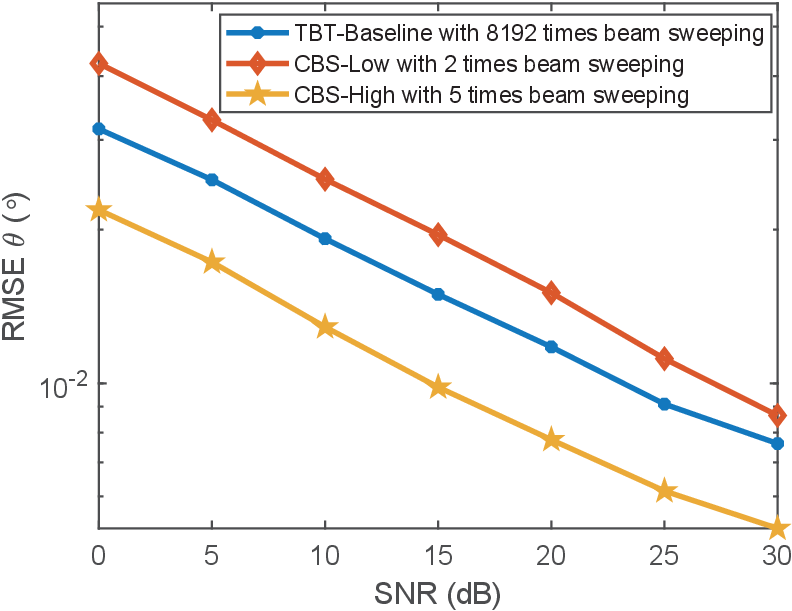}%
\label{fig_first_case}}
\hfil
\subfloat[]{\includegraphics[width=80mm]{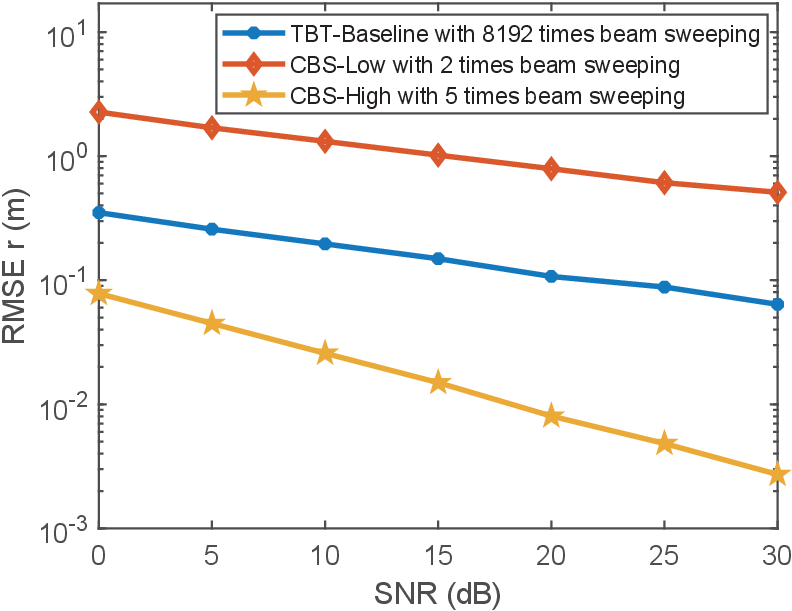}%
\label{fig_second_case}}
\caption{(a) Angle sensing RMSE of three single BS localization schemes.
(b) Distance sensing RMSE of three single BS localization schemes.
Note that different schemes require different beam sweeping times.}
\label{fig_sim}
\end{figure*}

\begin{figure*}[!t]
\centering
\subfloat[]{\includegraphics[width=80mm]{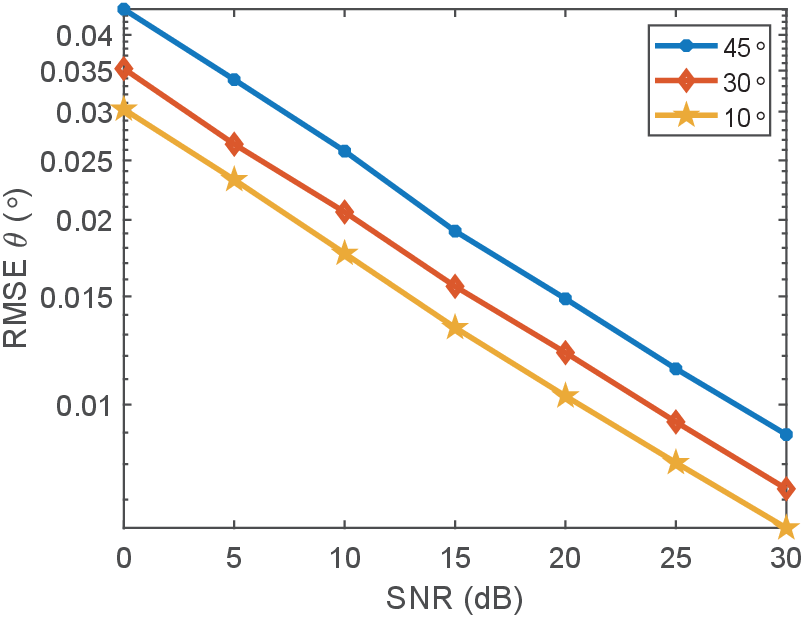}%
\label{fig_first_case}}
\hfil
\subfloat[]{\includegraphics[width=80mm]{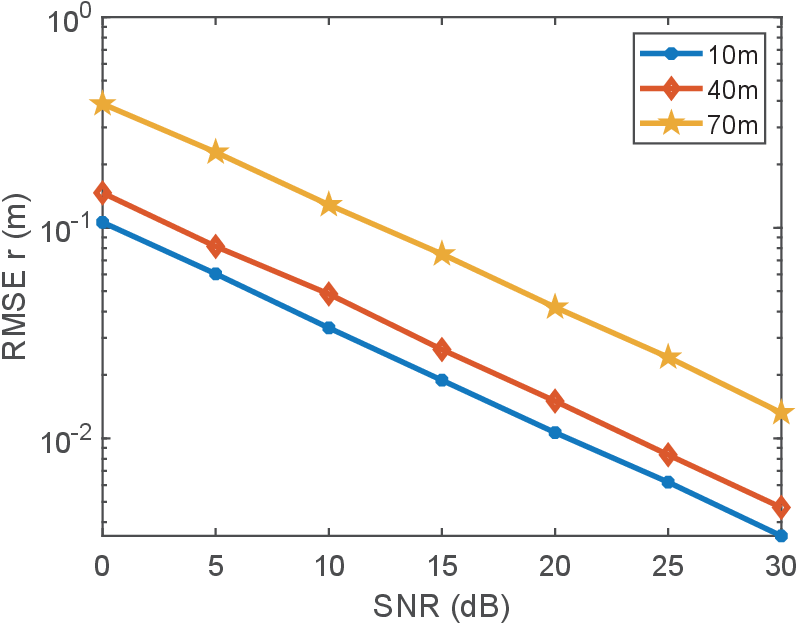}%
\label{fig_second_case}}
\caption{(a) Angle sensing RMSE of  the proposed CBS-High Scheme.
(b) Distance sensing RMSE of  the proposed CBS-High Scheme.
Different curves represent users at different locations.}
\label{fig_sim}
\end{figure*}

\subsection{Localization Performance Based on Single BS}

The TBT-Baseline  Scheme, proposed CBS-Low Scheme, and proposed CBS-High Scheme are the localization strategies based on single BS.
In the simulation of this subsection,
the number of BS antennas   is $N=256$, 
the  lowest carrier frequency is $f_0 = 60$ GHz, the OFDM  bandwidth  is $W = 4$~GHz,
 the number of subcarriers is $M+1=4096$ and the antenna spacing is $d=\frac{1}{2}\lambda = \frac{c}{2f_0}$.
Considering the Electromagnetic and Rayleigh distances, 
 the sensing range of BS system is set to
$\{(r,\theta)|5m \leq r \leq 50m, -60^\circ \leq \theta \leq 60^\circ\}$.
Then the average angle and distance intervals of controllable beam squint within the sensing range are $\frac{120^\circ}{4096-1}=0.029^\circ$ and $\frac{45m}{4096-1}=0.011m$, respectively.
To compare these  schemes more fairly, we set $I^a=M+1=4096$ and $I^d=M+1=4096$ in TBT-Baseline  Scheme.
This setting can ensure that the spatial scanning interval between TBT-Baseline Scheme  and controllable beam squint is basically consistent.

Fig.~11 shows the curves of the localization RMSE with SNR for three Schemes.
Under such parameter settings, the TBT-Baseline  Scheme requires $4096+4096=8192$ times beam sweeping 
 to complete the localization for the first user,
and   additional $4096$ times beam sweeping  are required for each additional user.
 The proposed CBS-Low Scheme only needs 2 times beam sweeping  to obtain the position of the first user, additional one time beam sweeping is required for each additional user, and its localization performance 
 is only slightly lower than the TBT-Baseline  Scheme.
This means that the CBS-Low Scheme can realize acceptable  user localization accuracy
while saves $99.9755\%$ of the beam sweeping cost compared to the TBT-Baseline  Scheme,
  thanks to  the wide coverage characteristics of  beam squint.
In addition, it can be seen that  the RMSE of the CBS-High Scheme  with only $P = 5$ times beam sweeping  
 is much lower than that of TBT-Baseline  Scheme  with $8192$ times  beam sweeping.
And the CBS-High Scheme does not require additional beam sweeping  for additional user localization.
This means that the beam sweeping  speed of the  CBS-High Scheme  is at least $1638$ times faster than that of TBT-Baseline  Scheme.
The simulation results indicate that   the  frequency-domain beam sweeping based on controllable beam squint can replace traditional time-domain beam sweeping, which can  greatly reduce the time cost of beam sweeping for sensing.

\subsection{Localization Performance of Users at Different Positions in CBS-High Scheme}
 Here, we will demonstrate the localization performance of the proposed CBS-High Scheme for users at different locations.
Assume the number of BS antennas  is $N=128$, the carrier frequency is $f_0=60$ GHz, the transmission bandwidth is $W = 1$ GHz, and the number of subcarriers is $M+1=2048$. The sensing range of BS system is set to
$\{(r,\theta)|5m \leq r \leq 80m, -60^\circ \leq \theta \leq 60^\circ\}$. Fig.~12 shows the RMSE  for locating users at different positions, in which
 both $\rm RMSE_{\theta}$ and $\rm RMSE_r$  decrease with the increase of SNR.

The three users in Fig.~12 (a) are at the same distance but different angles.
It can be found  that the closer the user's angle is to $0^\circ$, the lower the 
${\rm RMSE_\theta}$ will be.
This phenomenon is mainly due to the fact that  controllable beam squint gathers more subcarriers near $0^\circ$ in angle sensing stage,
which can be inferred from (19),
 resulting in better angle sensing performance near $0^\circ$.
Besides,
the three users in Fig.~12 (b) are at the same angle but different distances.
It shows that
the users who are  closer to the BS have higher distance sensing accuracy, 
which is due to the fact that near-field beamforming has better focusing performance at closer distances.

\subsection{The Impact of Beam Sweeping Times on  Localization Performance in  CBS-High Scheme}

\begin{figure*}[!t]
\centering
\subfloat[]{\includegraphics[width=80mm]{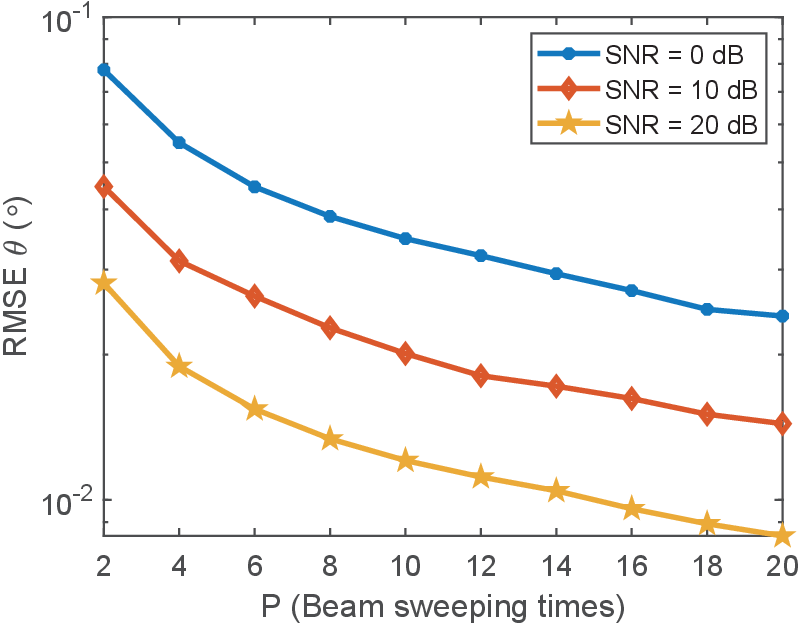}%
\label{fig_first_case}}
\hfil
\subfloat[]{\includegraphics[width=80mm]{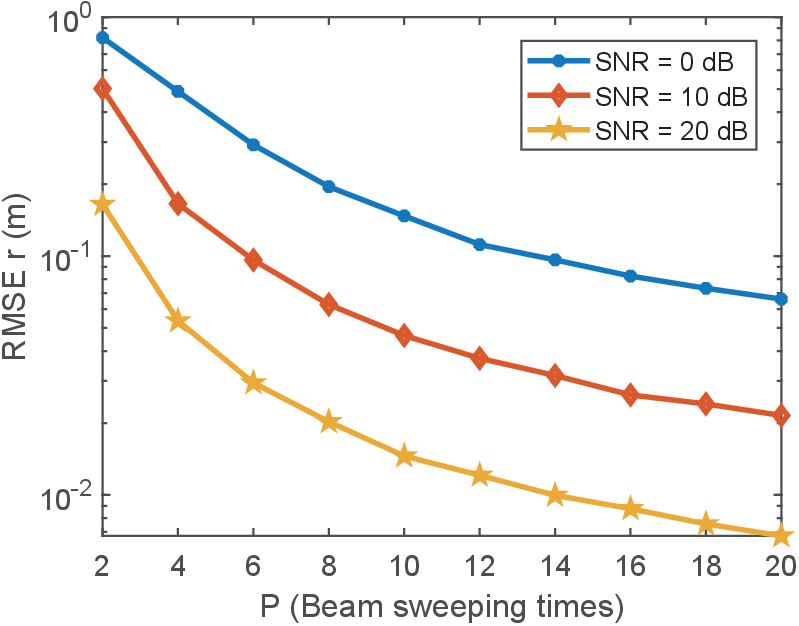}%
\label{fig_second_case}}
\caption{(a) The angle sensing RMSE of the proposed CBS-High Scheme with different $P$.
(b) The distance sensing RMSE of the proposed CBS-High Scheme with different $P$.}
\label{fig_sim}
\end{figure*}

 The CBS-High Scheme
 requires the implementation of $P$ times TTDs-assisted beamforming with different angle squint ranges  to realize user localization. 
Assuming the number of BS antennas is $N=128$, the number of subcarriers is $M+1=2048$ and the user is at $(60m, 20^\circ)$.
Fig.~13 shows the variation curve of the localization RMSE with the number of beam sweeping times $P$ under a fixed SNR.
It can be seen that both the $\rm RMSE_\theta$ and $\rm RMSE_r$  will decrease as the  increase of $P$. 
When $P = 4$ and SNR $= 10$ dB, the $\rm RMSE_\theta$ is $0.031^\circ$ and the $\rm RMSE_r$ is $0.165m$. When $P$ increases to 12, the $\rm RMSE_\theta$ decreases to $0.018^\circ$ and the $\rm RMSE_r$ decreases to $0.037m$. Besides, increasing $P$ also requires more time to complete sensing. Therefore, we can balance between the sensing accuracy and the acceptable beam sweeping time overhead.

\begin{figure}
\centering
\includegraphics[width=80mm]{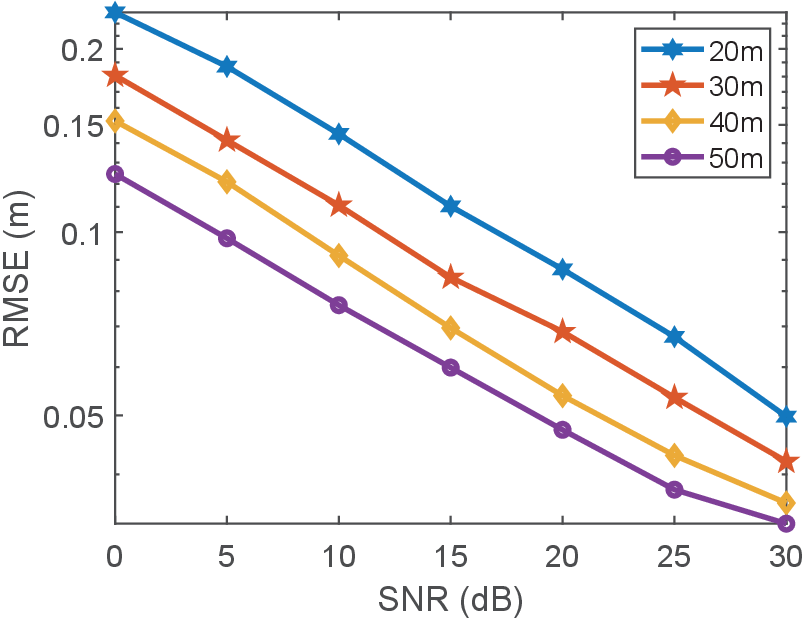}
\caption{The 2D localization RMSE of the proposed double BSs and beam squint based user localization scheme. The users' angles are $30^\circ$ and users' distances are different. $L=100m$ , $N=256$ and $M+1=2048$.}
\label{fig1:env}
\end{figure}

\subsection{Simulation of Double BSs and Beam Squint Based  Localization Scheme}

In the simulation of double BSs and beam squint based localization
 Scheme, the number of BS antennas is set as $N=256$, the number of subcarriers is $M+1=2048$, and the distance between the two BSs  is $L=100m$.
The 2D localization RMSE of the proposed CBS-2BS Scheme  is shown in Fig.~14.
It can be seen that the average 2D RMSE of all users is  approximately $0.17m$ when SNR is $0$ dB,  and gradually decrease to $0.10m$ when SNR increases to $15$ dB.
This indicates that the proposed CBS-2BS Scheme  can obtain the high-precision position estimations for all users with the cost of only two times beam sweeping.
In addition,  the 2D RMSE of the users whose abscissa ($x_{sp}$) is close to $\frac{L}{2}$ will be smaller, which is caused by geometric symmetry.

\section{Conclusion}

This paper proposes the low-overhead user localization schemes based on  beam squint in near-field ISAC system.
With the aid of the TTDs, the range and trajectory of the beam squint can be freely controlled, and hence it is possible to reversely utilize the beam squint for user localization.
Then we derive the trajectory equation for near-field beam squint points and design a way to control the trajectory of these beam squint points.
With the proposed design, beamforming from different subcarriers would purposely point to different angles and different distances such that users from different positions would receive the maximum power at different subcarriers. Hence, one can simply obtain the different users' position from the beam squint effect.  
Furthermore, we  utilize the phase difference of the maximum power subcarriers received by the user at different frequencies in several times beam sweeping to obtain a more accurate  distance estimation result, ultimately realizing high accuracy and low beam sweeping overhead user localization.
Simulation results demonstrate the effectiveness of the proposed schemes.

\bibliographystyle{ieeetr}
\bibliography{refersxsd.bib}

\begin{thebibliography}{10}

\bibitem{ISAC1}
C.~De~Lima, D.~Belot, R.~Berkvens, A.~Bourdoux, D.~Dardari, M.~Guillaud,
  M.~Isomursu, E.-S. Lohan, Y.~Miao, A.~N. Barreto, M.~R.~K. Aziz,
  J.~Saloranta, T.~Sanguanpuak, H.~Sarieddeen, G.~Seco-Granados, J.~Suutala,
  T.~Svensson, M.~Valkama, B.~Van~Liempd, and H.~Wymeersch, ``Convergent
  communication, sensing and localization in 6{G} systems: An overview of
  technologies, opportunities and challenges,'' {\em IEEE Access}, vol.~9,
  pp.~26902--26925, Jan. 2021.

\bibitem{9737357}
F.~Liu, Y.~Cui, C.~Masouros, J.~Xu, T.~X. Han, Y.~C. Eldar, and S.~Buzzi,
  ``Integrated sensing and communications: Toward dual-functional wireless
  networks for 6{G} and beyond,'' {\em IEEE J. Sel. Areas Commun.}, vol.~40,
  no.~6, pp.~1728--1767, Jun. 2022.

\bibitem{ISAC3}
D.~K. Pin~Tan, J.~He, Y.~Li, A.~Bayesteh, Y.~Chen, P.~Zhu, and W.~Tong,
  ``Integrated sensing and communication in 6{G}: Motivations, use cases,
  requirements, challenges and future directions,'' in {\em Proc. 1st IEEE Int.
  Online Symp. Joint Commun. Sens. (JCS)}, Dresden, Germany, Feb. 2021, pp.
  1--6.

\bibitem{ISAC2}
J.~Yang, C.-K. Wen, and S.~Jin, ``Hybrid active and passive sensing for {SLAM}
  in wireless communication systems,'' {\em IEEE J. Sel. Areas Commun.},
  vol.~40, no.~7, pp.~2146--2163, Mar. 2022.

\bibitem{d2}
G.~Kwon, A.~Conti, H.~Park, and M.~Z. Win, ``Joint communication and
  localization in millimeter wave networks,'' {\em IEEE J. Sel. Topics Signal
  Process.}, vol.~15, no.~6, pp.~1439--1454, Nov. 2021.

\bibitem{butalso1}
Q.~Zhang, H.~Sun, Z.~Wei, and Z.~Feng, ``Sensing and communication integrated
  system for autonomous driving vehicles,'' in {\em Proc. IEEE INFOCOM WKSHPS},
  Toronto, ON, Canada, Jul. 2020, pp. 1278--1279.

\bibitem{jiankong}
A.~{Bejarano-Carbo}, H.~{An}, K.~{Choo}, S.~{Liu}, Q.~{Zhang}, D.~{Sylvester},
  D.~{Blaauw}, and H.-S. {Kim}, ``{Millimeter-scale ultra-low-power imaging
  system for intelligent edge monitoring},'' {\em arXiv e-prints},
  p.~arXiv:2203.04496, Mar. 2022.

\bibitem{UAV1}
X.~{Jing}, F.~{Liu}, C.~{Masouros}, and Y.~{Zeng}, ``{ISAC from the sky: {UAV}
  trajectory design for joint communication and target localization},'' {\em
  arXiv e-prints}, p.~arXiv:2207.02904, Jul. 2022.

\bibitem{music1}
R.~O. Schmidt, ``Multiple emitter location and signal parameter estimation,''
  {\em IEEE Trans. Antennas Propag.}, vol.~34, no.~3, pp.~276--280, Mar. 1986.

\bibitem{cs1}
Q.~Shen, W.~Liu, W.~Cui, and S.~Wu, ``Underdetermined {DOA} estimation under
  the compressive sensing framework: A review,'' {\em IEEE Access}, vol.~4,
  pp.~8865--8878, Nov. 2016.

\bibitem{cs2}
F.~Yan, M.~Jin, and X.~Qiao, ``Low-complexity {DOA} estimation based on
  compressed {MUSIC} and its performance analysis,'' {\em IEEE Trans. Signal
  Process.}, vol.~61, no.~8, pp.~1915--1930, Apr. 2013.

\bibitem{cs3}
S.~Fortunati, R.~Grasso, F.~Gini, and M.~S. Greco, ``Single snapshot {DOA}
  estimation using compressed sensing,'' in {\em Proc. IEEE Int. Conf. Acoust.
  Speech Signal Process.}, Florence, Italy, Jul. 2014, pp. 2297-2301.

\bibitem{DOAISACPZ}
Z.~{Chen}, P.~{Chen}, Z.~{Guo}, and X.~{Wang}, ``{A {RIS}-based passive {DOA}
  estimation method for integrated sensing and communication system},'' {\em
  arXiv e-prints}, p.~arXiv:2204.11626, Apr. 2022.

\bibitem{DOAISACPC}
P.~{Chen}, Z.~{Chen}, L.~{Liu}, Y.~{Chen}, and X.~{Wang}, ``{{SDOA}net: An
  efficient deep learning-based {DOA} estimation network for imperfect
  array},'' {\em arXiv e-prints}, p.~arXiv:2203.10231, Mar. 2022.

\bibitem{ruili}
K.~R. Dandekar, ``Reviews and abstracts [review of wireless communications by
  {Andreas F. Molisch}; 2005],'' {\em IEEE Antenn. Propag. Mag.}, vol.~49,
  no.~1, pp.~132--133, Feb. 2007.

\bibitem{zhy11}
H.~{Zhang}, N.~{Shlezinger}, F.~{Guidi}, D.~{Dardari}, and Y.~C. {Eldar},
  ``{6{G} wireless communications: From far-field beam steering to near-field
  beam focusing},'' {\em arXiv e-prints}, p.~arXiv:2203.13035, Mar. 2022.

\bibitem{thz1112}
A.~M. {Elbir}, K.~{Vijay Mishra}, S.~{Chatzinotas}, and M.~{Bennis},
  ``{Terahertz-band integrated sensing and communications: Challenges and
  opportunities},'' {\em arXiv e-prints}, p.~arXiv:2208.01235, Aug. 2022.

\bibitem{near1}
H.~Zhang, N.~Shlezinger, F.~Guidi, D.~Dardari, M.~F. Imani, and Y.~C. Eldar,
  ``Near-field wireless power transfer for 6{G} internet of everything mobile
  networks: Opportunities and challenges,'' {\em IEEE Commun. Mag.}, vol.~60,
  no.~3, pp.~12--18, Mar. 2022.

\bibitem{9941256}
Y.~Jiang, F.~Gao, M.~Jian, S.~Zhang, and W.~Zhang, ``Reconfigurable intelligent
  surface for near field communications: Beamforming and sensing,'' {\em IEEE
  Trans. Wireless Commun.,}, pp.~1--1, Nov. 2022.

\bibitem{yang1}
J.~Yang, Y.~Zeng, S.~Jin, C.-K. Wen, and P.~Xu, ``Communication and
  localization with extremely large lens antenna array,'' {\em IEEE Trans.
  Wireless Commun.}, vol.~20, no.~5, pp.~3031--3048, May 2021.

\bibitem{sec1duan1cite1}
H.~Sarieddeen, N.~Saeed, T.~Y. Al-Naffouri, and M.-S. Alouini, ``Next
  generation terahertz communications: A rendezvous of sensing, imaging, and
  localization,'' {\em IEEE Commun. Mag.}, vol.~58, no.~5, pp.~69--75, May
  2020.

\bibitem{sec1duan1cite2}
Z.~Zhang, Y.~Xiao, Z.~Ma, M.~Xiao, Z.~Ding, X.~Lei, G.~K. Karagiannidis, and
  P.~Fan, ``6{G} wireless networks: Vision, requirements, architecture, and key
  technologies,'' {\em IEEE Veh. Technol. Mag.}, vol.~14, no.~3, pp.~28--41,
  Sep. 2019.

\bibitem{sec1duan1cite3}
E.~A. Kadir, R.~Shubair, S.~K. Abdul~Rahim, M.~Himdi, M.~R. Kamarudin, and
  S.~L. Rosa, ``B5{G} and {6G}: Next generation wireless communications
  technologies, demand and challenges,'' in {\em Proc. Int. Congr. Adv.
  Technol. Eng. (ICOTEN)}, Taiz, Yemen, Jul. 2021, pp. 1--6.

\bibitem{wblsp}
B.~Wang, F.~Gao, S.~Jin, H.~Lin, and G.~Y. Li, ``Spatial- and
  frequency-wideband effects in millimeter-wave massive {MIMO} systems,'' {\em
  IEEE Trans. Signal Process.}, vol.~66, no.~13, pp.~3393--3406, Jul. 2018.

\bibitem{b1}
K.~Spoof, V.~Unnikrishnan, M.~Zahra, K.~Stadius, M.~Kosunen, and J.~Ryynänen,
  ``True-time-delay beamforming receiver with {RF} re-sampling,'' {\em IEEE
  Trans. Circuits Syst. I: Regular Papers}, vol.~67, no.~12, pp.~4457--4469,
  Dec. 2020.

\bibitem{9896734}
Y.~Wu, G.~Song, H.~Liu, L.~Xiao, and T.~Jiang, ``3-{D} hybrid beamforming for
  terahertz broadband communication system with beam squint,'' {\em IEEE Trans.
  Broadcast.}, vol.~69, no.~1, pp.~264--275, Mar. 2023.

\bibitem{9839132}
D.~Q. Nguyen and T.~Kim, ``Joint delay and phase precoding under true-time
  delay constraints for {THz} massive {MIMO},'' in {\em Proc. IEEE Int. Conf.
  Commun. (ICC)}, Seoul, Korea, May 2022, pp. 3496--3501.

\bibitem{nn1}
F.~Gao, B.~Wang, C.~Xing, J.~An, and G.~Y. Li, ``Wideband beamforming for
  hybrid massive {MIMO} terahertz communications,'' {\em IEEE J. Sel. Areas
  Commun.}, vol.~39, no.~6, pp.~1725--1740, Jun. 2021.

\bibitem{10058989}
F.~Gao, L.~Xu, and S.~Ma, ``Integrated sensing and communications with joint
  beam-squint and beam-split for {mmWave}/{THz} massive {MIMO},'' {\em IEEE
  Trans. Commun.}, pp.~1--1, Mar. 2023.

\bibitem{waveassume1}
R.~Kumaresan and D.~W. Tufts, ``Estimating the angles of arrival of multiple
  plane waves,'' {\em IEEE Trans. Aerosp. Electron. Syst.}, vol.~AES-19, no.~1,
  pp.~134--139, Jan. 1983.

\bibitem{waveassume2}
A.~Broquetas, J.~Palau, L.~Jofre, and A.~Cardama, ``Spherical wave near-field
  imaging and radar cross-section measurement,'' {\em IEEE Trans. Antennas
  Propag.}, vol.~46, no.~5, pp.~730--735, May 1998.

\bibitem{waveassume3}
A.~Ludwig, ``Near-field far-field transformations using spherical-wave
  expansions,'' {\em IEEE Trans. Antennas Propag.}, vol.~19, no.~2,
  pp.~214--220, Mar. 1971.

\bibitem{jinsi}
J.-W. Tao, L.~Liu, and Z.-Y. Lin, ``Joint {DOA}, range, and polarization
  estimation in the {Fresnel} region,'' {\em IEEE Trans. Aerosp. Electron.
  Syst.}, vol.~47, no.~4, pp.~2657--2672, Oct. 2011.

\bibitem{xindao}
Z.~Zhou, X.~Gao, J.~Fang, and Z.~Chen, ``Spherical wave channel and analysis
  for large linear array in {LoS} conditions,'' in {\em Proc. IEEE Globecom
  Workshops}, San Diego, CA, USA, Dec. 2015, pp. 1--6.

\bibitem{7394105}
R.~Rotman, M.~Tur, and L.~Yaron, ``True time delay in phased arrays,'' {\em
  Proceedings of the IEEE}, vol.~104, no.~3, pp.~504--518, Mar. 2016.

\bibitem{radarTTDs2}
M.~Schartel, W.~Mayer, and C.~Waldschmidt, ``Digital true time delay for pulse
  correlation radars,'' in {\em Proc. 46th Eur. Microw. Conf. (EuMC)}, London,
  UK, Oct. 2016, pp. 1477--1480.

\bibitem{2022arXiv220602996S}
Y.~{Sun}, J.~{Li}, T.~{Zhang}, R.~{Wang}, X.~{Peng}, T.~{Xiao Han}, and
  H.~{Tan}, ``{An indoor environment sensing and localization system via mmWave
  phased array},'' {\em arXiv e-prints}, p.~arXiv:2206.02996, Jun. 2022.

\bibitem{DBLP23}
T.~Wei, A.~Zhou, and X.~Zhang, ``Facilitating robust 60 {GHz} network
  deployment by sensing ambient reflectors,'' in {\em Proc. 14th USENIX Symp.
  Network. Systems Design and Implementation}, USENIX Association, BOSTON, MA,
  Mar. 2017, pp. 213--226.

\end{thebibliography}

\begin{IEEEbiography}[{\includegraphics[width=1in,height=1.25in,clip,keepaspectratio]{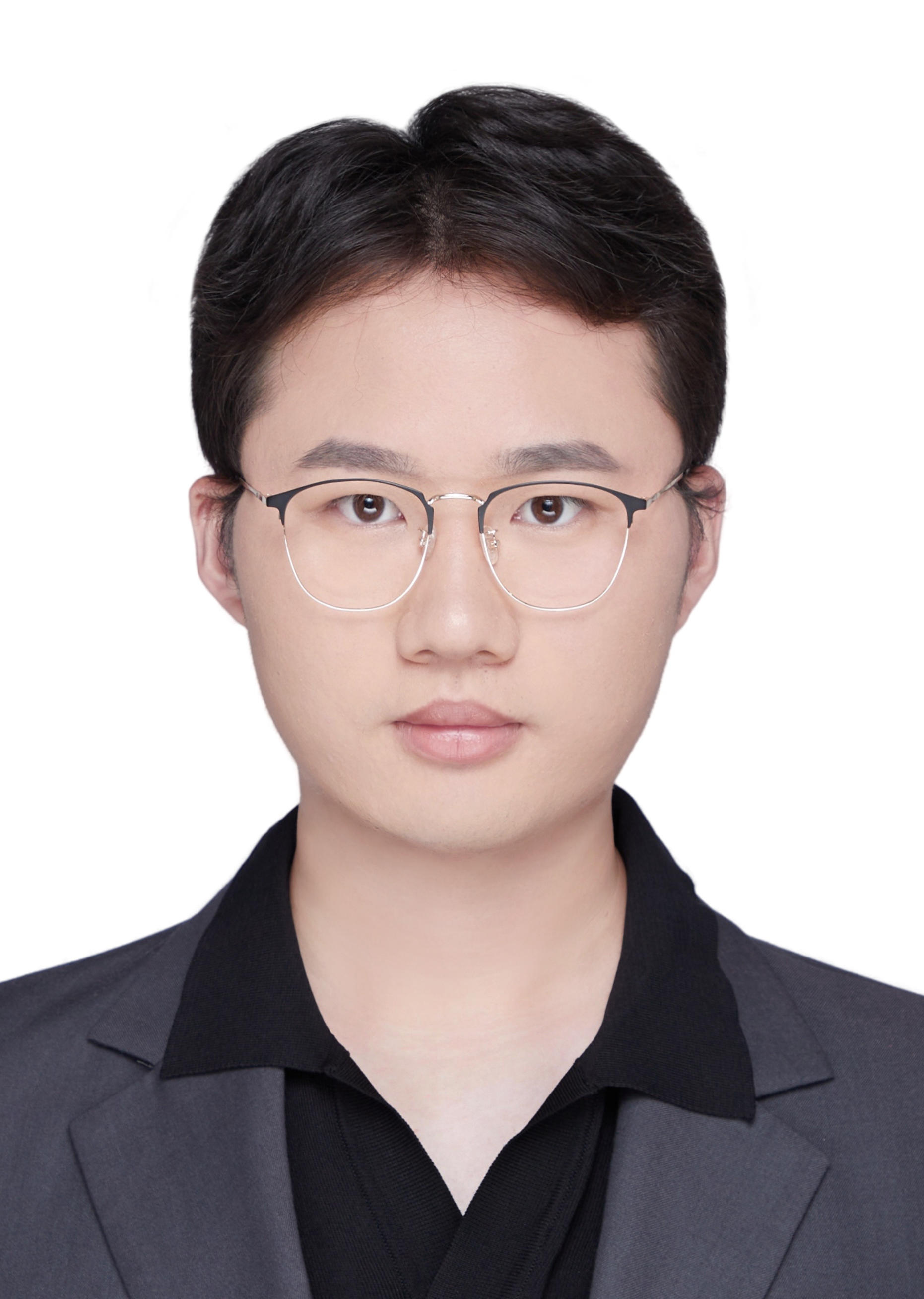}}]{\\Hongliang Luo}
received the B.Eng. degree from Xidian University, Xi'an, China, in 2023.
He is currently working toward the Ph.D. degree with the Department of Automation, Tsinghua University, Beijing, China.

His research interests include wireless communication, radar sensing, array signal processing, massive MIMO and beamforimg design.
\end{IEEEbiography}

\begin{IEEEbiography}[{\includegraphics[width=1in,height=1.25in,clip,keepaspectratio]{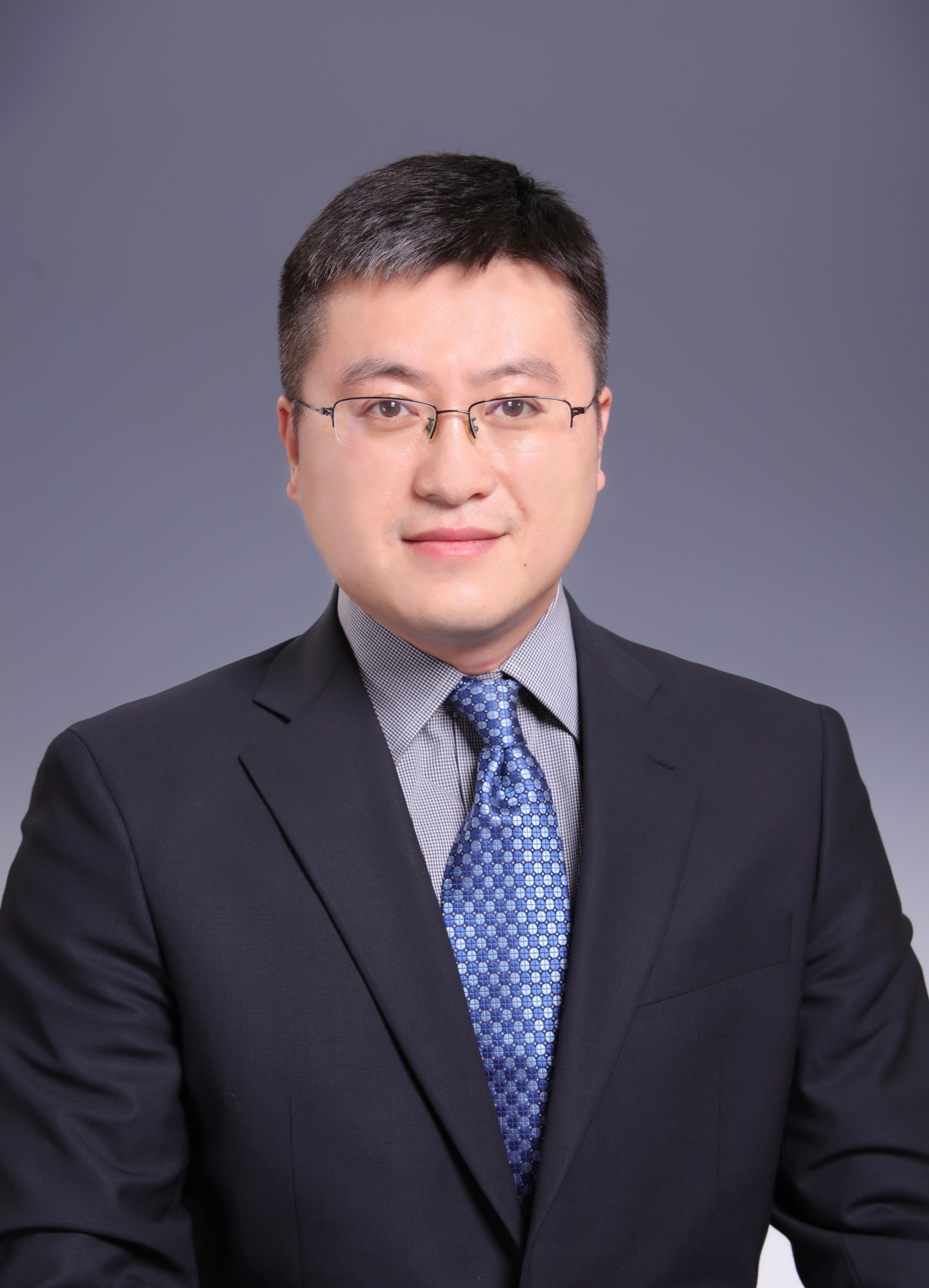}}]{Feifei Gao} 
(Fellow, IEEE)
received the B.Eng. degree from Xi'an Jiaotong University, Xi'an, China in 2002, the M.Sc. degree from McMaster University, Hamilton, ON, Canada in 2004, and the Ph.D. degree from National University of Singapore, Singapore in 2007. Since 2011, he joined the Department of Automation, Tsinghua University, Beijing, China, where he is currently an Associate Professor.

Prof. Gao's research interests include signal processing for communications, array signal processing, convex optimizations, and artificial intelligence assisted communications. He has authored/coauthored more than 200 refereed IEEE journal papers and more than 150 IEEE conference proceeding papers that are cited more than 16000 times in Google Scholar. Prof. Gao has served as an Editor of IEEE Transactions on Wireless Communications, IEEE Journal of Selected Topics in Signal Processing (Lead Guest Editor), IEEE Transactions on Cognitive Communications and Networking, IEEE Signal Processing Letters (Senior Editor), IEEE Communications Letters (Senior Editor), IEEE Wireless Communications Letters, and China Communications. He has also served as the symposium co-chair for 2019 IEEE Conference on Communications (ICC), 2018 IEEE Vehicular Technology Conference Spring (VTC), 2015 IEEE Conference on Communications (ICC), 2014 IEEE Global Communications Conference (GLOBECOM), 2014 IEEE Vehicular Technology Conference Fall (VTC), as well as Technical Committee Members for more than 50 IEEE conferences.
\end{IEEEbiography}

\begin{IEEEbiography}[{\includegraphics[width=1in,height=1.25in,clip,keepaspectratio]{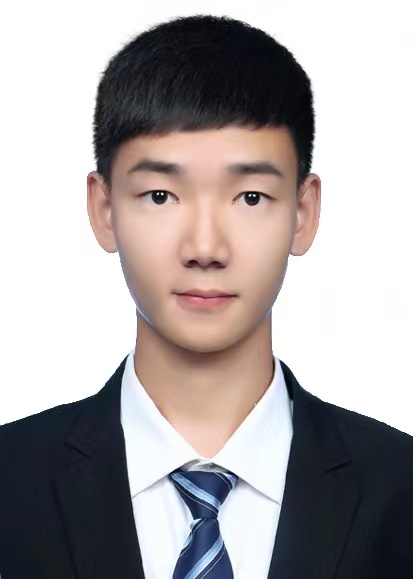}}]{Wanmai Yuan}
(Student Member, IEEE)
is a senior engineer in the Information Science Academy of CETC, Beijing, China. He received the B.Eng. degree in communication engineering from Xidian University, Xi'an, China, in June 2014, and the Ph.D. degree in electronics and information engineering from Harbin Institute of Technology, China in July 2019. He also received the Ph.D. degree in electronic and information engineering, the Hong Kong Polytechnic University, Hong Kong in Sep 2019. From 2018 to 2019, he was a Ph.D. visiting student in the Department of Electrical and Computer Engineering, University of Toronto. 

His main research interests include flocking control and formation control for UAVs. He received Young Elite Scientist Sponsorship by China Association for Science and Technology (CAST).
\end{IEEEbiography}

\begin{IEEEbiography}[{\includegraphics[width=1in,height=1.2in,clip,keepaspectratio]{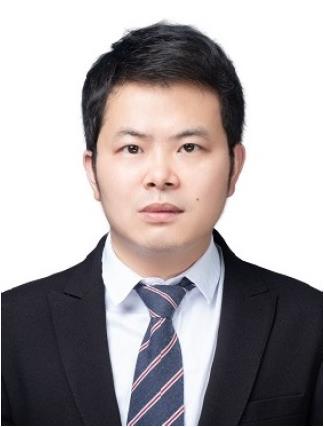}}]
{Shun Zhang}
(Senior Member, IEEE)  received the B.S. degree in communication engineering from Shandong University, Jinan, China, in 2007, and the Ph.D. degree in communications and signal processing from Xidian University, Xi’an, China, in 2013. He is currently with the State Key Laboratory of Integrated Services Networks, Xidian University, where he is currently a Professor.

Prof. Zhang's research interests include massive MIMO, millimeter wave systems, RIS assisted communications, deep learning for communication systems, orthogonal time frequency space (OTFS) systems, and multiple access techniques. He is an Editor for Physical Communication. He has authored or coauthored more than 80 journal and conference papers, and is the inventor of 16 granted patents (including a PCT patent authorized by US Patent and Trademark Office). He has received two Best Paper Awards in conferences, and two prize awards in natural sciences for research excellence by both China Institute of Communications and Chinese Institute of Electronics.
\end{IEEEbiography}

\vfill

\end{document}